\documentclass{ar-1col-S2O}
\usepackage[dvipsnames]{xcolor}
\usepackage{xspace}
\usepackage[sort&compress,numbers]{natbib}
\usepackage{url}
\usepackage{derivative}
\usepackage[hyperfootnotes=false]{hyperref}
\usepackage{tikz}
\usepackage{amsmath,amssymb}
\usepackage{booktabs}
\jname{Ann. Rev. Nucl. Part. Sci.}
\jvol{76}
\jyear{2026}
\doi{10.1146/annurev-nucl-102422-040841}

\begin{document}

\newcommand{\muvec}{\boldsymbol{\mu}}
\newcommand{\Svec}{{\mathbf S}}
\newcommand{\vecB}{{\mathbf B}}
\newcommand{\vecP}{{\mathbf P}}
\newcommand{\vecE}{{\mathbf E}}
\newcommand{\vecomega}{{\boldsymbol{\omega}}}
\newcommand{\vecbeta}{{\boldsymbol{\beta}}}
\newcommand{\Trprime}{\ensuremath{T'_{r}}\xspace}
\newcommand{\Tr}{\ensuremath{T^{}_{r}}\xspace}
\newcommand{\opprime}{\ensuremath{\omega'^{}_p}\xspace}
\newcommand{\opprimetildeatTexp}{\ensuremath{\tilde{\omega}'^{}_p(\Tr)}}
\newcommand{\wa}{\ensuremath{\omega_{a}}\xspace}
\newcommand{\wam}{\ensuremath{\omega_{a}^{m}}\xspace}
\newcommand{\Rmu}{\ensuremath{{\mathcal R}_\mu}\xspace}
\newcommand{\Rmuprime}{\ensuremath{{\mathcal R}'^{}_\mu}\xspace}

\newcommand{\amu}{\ensuremath{a_{\mu}}}
\newcommand{\amuSM}{\ensuremath{a_\mu^\text{SM}}}
\newcommand{\amuexp}{\ensuremath{a_\mu^\text{exp}}}
\newcommand{\amuHVP}{\ensuremath{a_\mu^\text{HVP}}}
\newcommand{\amuHVPLO}{\ensuremath{a_\mu^\text{HVP, LO}}}
\newcommand{\amuHVPNLO}{\ensuremath{a_\mu^\text{HVP, NLO}}}
\newcommand{\amuHVPNNLO}{\ensuremath{a_\mu^\text{HVP, NNLO}}}
\newcommand{\amuHLbL}{\ensuremath{a_\mu^\text{HLbL}}}
\newcommand{\amuHLbLNLO}{\ensuremath{a_\mu^\text{HLbL, NLO}}}
\newcommand{\amuhad}{\ensuremath{a_\mu^\text{had}}}
\newcommand{\amuQED}{\ensuremath{a_\mu^\text{QED}}}
\newcommand{\amuEW}{\ensuremath{a_\mu^\text{EW}}}
\newcommand{\amuwin}{\ensuremath{a_\mu^\text{win}}}

\newcommand{\babar}{\textsc{BaBar}}
\newcommand{\Order}{{\mathcal O}}

%key numbers from WP25
\newcommand{\amuexpresult}{116\, 592\, 071.5(14.5)}

\newcommand{\amuHVPLOresult}{7132(61)}
\newcommand{\amuHVPNLOresult}{-99.6(1.3)}
\newcommand{\amuHVPNNLOresult}{12.4(1)}
\newcommand{\amuHVPtotalresult}{7045(61)}

\newcommand{\amuHLbLdataresult}{103.3(8.8)}
\newcommand{\amuHLbLlatticeresult}{122.5(9.0)}
\newcommand{\amuHLbLNLOdataresult}{2.6(6)}
\newcommand{\amuHLbLaverageresult}{112.6(9.6)}
\newcommand{\amuHLbLtotalresult}{115.5(9.9)}

\newcommand{\amuQEDresult}{116\,584\,718.8(2)}
\newcommand{\amuEWresult}{154.4(4)}

\newcommand{\amuSMresult}{116\,592\,033(62)}
\newcommand{\amudiffresult}{38(63)}
%%%%%%%%%%%%%%%%%
%%%Added by Dave from PRD

%\newcommand{\amu}{\ensuremath{a^{}_{\mu}}\xspace}
%\newcommand{\amuSM}{\ensuremath{a^{SM}_{\mu}}\xspace}
\newcommand{\amuEXP}{\ensuremath{a^{EXP}_{\mu}}\xspace}
\newcommand{\amuHLO}{\ensuremath{a^{HLO}_{\mu}}\xspace}
\newcommand{\gm}{\ensuremath{g-2}\xspace}
\newcommand{\gmtwo}{\gm}
\newcommand{\oa}{\ensuremath{\omega^{}_a}\xspace}
\newcommand{\oam}{\ensuremath{\omega^{m}_a}\xspace}
\newcommand{\ocycl}{\ensuremath{\omega^{}_c}\xspace}
\newcommand{\oS}{\ensuremath{\omega^{}_S}\xspace}
\newcommand{\op}{\ensuremath{\omega^{}_p}\xspace}
\newcommand{\opmeas}{\omega_{p}^{\pp}\xspace}
\newcommand{\opprimeatTexp}{\ensuremath{\omega'^{}_p(\Tr)}\xspace}
\newcommand{\opprimetilde}{\ensuremath{\tilde{\omega}'^{}_p}\xspace}
\newcommand{\opprimetildeofT}{\ensuremath{\tilde{\omega}'^{}_p(T)}\xspace}
\newcommand{\optilde}{\ensuremath{\tilde{\omega}^{}_p}\xspace}
\newcommand{\omegap}{\op}
\newcommand{\omegapfree}{\ensuremath{\omega_p^\mathrm{free}}\xspace}
\newcommand{\fid}{FID\xspace}

\newcommand{\beq}{\begin{equation}}
\newcommand{\eeq}{\end{equation}}
\newcommand{\keV}{\,\text{keV}}
\newcommand{\MeV}{\,\text{MeV}}
\newcommand{\GeV}{\,\text{GeV}}
\newcommand{\TeV}{\,\text{TeV}}
\newcommand{\ppb}{\,\text{ppb}}
\newcommand{\ppm}{\,\text{ppm}}

%citation lists

\newcommand{\expref}{Muong-2:2025xyk,Muong-2:2023cdq,Muong-2:2024hpx,Muong-2:2021ojo,Muong-2:2021vma,Muong-2:2021ovs,Muong-2:2021xzz,Muong-2:2006rrc}

\newcommand{\QEDref}{Aoyama:2012wk,Volkov:2019phy,Volkov:2024yzc,Aoyama:2024aly,Parker:2018vye,Morel:2020dww,Fan:2022eto}

\newcommand{\EWref}{Czarnecki:2002nt,Gnendiger:2013pva,Ludtke:2024ase,Hoferichter:2025yih}

\newcommand{\latticeHVPref}{RBC:2018dos,Giusti:2019xct,Borsanyi:2020mff,Lehner:2020crt,Wang:2022lkq,Aubin:2022hgm,Ce:2022kxy,ExtendedTwistedMass:2022jpw,RBC:2023pvn,Kuberski:2024bcj,Boccaletti:2024guq,Spiegel:2024dec,RBC:2024fic,Djukanovic:2024cmq,ExtendedTwistedMass:2024nyi,MILC:2024ryz,FermilabLatticeHPQCD:2024ppc}

\newcommand{\latticeHLbLref}{Blum:2019ugy,Chao:2021tvp,Chao:2022xzg,Blum:2023vlm,Fodor:2024jyn}

\newcommand{\dataHLbLref}{Colangelo:2015ama,Masjuan:2017tvw,Colangelo:2017fiz,Hoferichter:2018kwz,Eichmann:2019tjk,Bijnens:2019ghy,Leutgeb:2019gbz,Cappiello:2019hwh,Masjuan:2020jsf,Bijnens:2020xnl,Bijnens:2021jqo,Danilkin:2021icn,Stamen:2022uqh,Leutgeb:2022lqw,Hoferichter:2023tgp,Hoferichter:2024fsj,Estrada:2024cfy,Ludtke:2024ase,Deineka:2024mzt,Eichmann:2024glq,Bijnens:2024jgh,Hoferichter:2024bae,Holz:2024diw,Cappiello:2025fyf}

\newcommand{\HVPref}{RBC:2018dos,Giusti:2019xct,Borsanyi:2020mff,Lehner:2020crt,Wang:2022lkq,Aubin:2022hgm,Ce:2022kxy,ExtendedTwistedMass:2022jpw,RBC:2023pvn,Kuberski:2024bcj,Boccaletti:2024guq,Spiegel:2024dec,RBC:2024fic,Djukanovic:2024cmq,ExtendedTwistedMass:2024nyi,MILC:2024ryz,FermilabLatticeHPQCD:2024ppc,Keshavarzi:2019abf,DiLuzio:2024sps,Kurz:2014wya}

\newcommand{\HLbLref}{Colangelo:2015ama,Masjuan:2017tvw,Colangelo:2017fiz,Hoferichter:2018kwz,Eichmann:2019tjk,Bijnens:2019ghy,Leutgeb:2019gbz,Cappiello:2019hwh,Masjuan:2020jsf,Bijnens:2020xnl,Bijnens:2021jqo,Danilkin:2021icn,Stamen:2022uqh,Leutgeb:2022lqw,Hoferichter:2023tgp,Hoferichter:2024fsj,Estrada:2024cfy,Ludtke:2024ase,Deineka:2024mzt,Eichmann:2024glq,Bijnens:2024jgh,Hoferichter:2024bae,Holz:2024diw,Cappiello:2025fyf,Colangelo:2014qya,Blum:2019ugy,Chao:2021tvp,Chao:2022xzg,Blum:2023vlm,Fodor:2024jyn}

\newcommand{\SMref}{Aoyama:2012wk,Volkov:2019phy,Volkov:2024yzc,Aoyama:2024aly,Parker:2018vye,Morel:2020dww,Fan:2022eto,Czarnecki:2002nt,Gnendiger:2013pva,Ludtke:2024ase,Hoferichter:2025yih,RBC:2018dos,Giusti:2019xct,Borsanyi:2020mff,Lehner:2020crt,Wang:2022lkq,Aubin:2022hgm,Ce:2022kxy,ExtendedTwistedMass:2022jpw,RBC:2023pvn,Kuberski:2024bcj,Boccaletti:2024guq,Spiegel:2024dec,RBC:2024fic,Djukanovic:2024cmq,ExtendedTwistedMass:2024nyi,MILC:2024ryz,FermilabLatticeHPQCD:2024ppc,Keshavarzi:2019abf,DiLuzio:2024sps,Kurz:2014wya,Colangelo:2015ama,Masjuan:2017tvw,Colangelo:2017fiz,Hoferichter:2018kwz,Eichmann:2019tjk,Bijnens:2019ghy,Leutgeb:2019gbz,Cappiello:2019hwh,Masjuan:2020jsf,Bijnens:2020xnl,Bijnens:2021jqo,Danilkin:2021icn,Stamen:2022uqh,Leutgeb:2022lqw,Hoferichter:2023tgp,Hoferichter:2024fsj,Estrada:2024cfy,Deineka:2024mzt,Eichmann:2024glq,Bijnens:2024jgh,Hoferichter:2024bae,Holz:2024diw,Cappiello:2025fyf,Colangelo:2014qya,Blum:2019ugy,Chao:2021tvp,Chao:2022xzg,Blum:2023vlm,Fodor:2024jyn}

% Page header
\markboth{D.~W.~Hertzog and M.~Hoferichter}{$(g_\mu-2)$: status and perspectives}

% Title
\title{The anomalous magnetic moment of the muon: status and perspectives}

%Authors, affiliations address.
\author{David W. Hertzog$^1$ and Martin Hoferichter$^2$
\affil{$^1$University of Washington, Department of Physics, Box 351560, Seattle, WA 98195, USA}
\affil{$^2$Albert Einstein Center for Fundamental Physics, Institute for Theoretical Physics, University of Bern, Sidlerstrasse 5,
3012 Bern, Switzerland}
}

%Abstract
\begin{abstract}
We review the status of the anomalous magnetic moment of the muon as a precision probe of physics beyond the Standard Model (SM) after the release of the final results from the Fermi National Accelerator Laboratory (FNAL) Muon $g-2$ experiment and the second White Paper of the Muon $g-2$ Theory Initiative. While the SM prediction requires further improvements by a factor of four to fully leverage the sensitivity achieved in experiment, the FNAL measurement will set the standard for many years to come, and we discuss a variety of features of the experimental campaign that made this achievement possible. In going forward, we discuss current efforts to improve the SM prediction, and imagine how an experiment would have to be devised to surpass $124\ppb$ in precision.
\end{abstract}

%Keywords, etc.
\maketitle

%Table of Contents
\tableofcontents

\section{INTRODUCTION}
\label{sec:intro}

The anomalous magnetic moment of leptons, $a_\ell$, defines an observable that allows for stringent tests of the Standard Model (SM) of particle physics down to its quantum nature, provided that both measurement and theory prediction can be carried out to high precision. In particular, $a_\ell$ is a flavor- and $CP$-conserving quantity, deriving from quantum fluctuations within the SM beyond the Dirac value of the $g$ factor, $a_\ell=(g_\ell-2)/2$, while necessitating a mechanism for a chirality flip to connect left- and right-handed lepton fields. Accordingly, a generic heavy contribution beyond the SM (BSM) scales as~\cite{Athron:2025ets}
\beq
\label{naive_scaling}
a_\ell^\text{BSM}\simeq R_\chi \frac{c_L^\ell c_R^\ell}{16\pi^2}\times \frac{m_\ell^2}{\Lambda_\text{BSM}^2}\simeq \big(c_L^\ell c_R^\ell R_\chi\big) \bigg(\frac{1\TeV}{\Lambda_\text{BSM}}\bigg)^2\times\begin{cases}
    1.7\times 10^{-15}\qquad &\ell=e\,,\\
                 7.1\times 10^{-11}\qquad  &\ell=\mu\,,\\
                 2.0\times 10^{-8}\qquad  &\ell=\tau\,.
                 \end{cases}
\eeq
Setting the prefactor on the right-hand side equal to $1$ provides a useful benchmark for
evaluating the power of an experimental constraint. In Eq.~\eqref{naive_scaling}
the $c_{L,R}^\ell$ are $\Order(1)$ couplings (with the superscript indicating that the BSM physics need not be lepton flavor universal), $\Lambda_\text{BSM}$ refers to the BSM scale, the factor $16\pi^2$ is generic for a quantum-loop correction,
the $m_\ell^2$
dependence accounts for the chirality flip required in the magnetic moment operator, and $R_\chi$
is a factor to change the value if the chirality flip has a different origin. In specific models,
$R_\chi$ can induce so-called chiral enhancements~\cite{Athron:2025ets,Crivellin:2021rbq}, e.g., when the chirality flip is mediated by a new heavy fermion $F$, $R_\chi=m_F/m_\ell$, while another special case, that of radiative mass generation, would amount to $c_L^\ell c_R^\ell R_\chi\simeq 16\pi^2\simeq 160$. Further scenarios yielding contributions well above the benchmark~\eqref{naive_scaling} can be obtained in the presence of light new particles~\cite{Athron:2025ets}.

In the case of the electron, the comparison of direct measurement~\cite{Fan:2022eto} and SM prediction~\cite{Laporta:2017okg,Volkov:2019phy,Volkov:2024yzc,Aoyama:2024aly,Keshavarzi:2019abf,DiLuzio:2024sps,Jegerlehner:2017zsb,Hoferichter:2025fea} constrains BSM effects 
to $|a_e^\text{BSM}|\lesssim 1.0\times 10^{-12}$.  However, at present the sensitivity is limited by a $5.5\sigma$ tension in the fine-structure constant $\alpha$~\cite{Parker:2018vye,Morel:2020dww}, whose resolution would propel the reach to $1.5\times 10^{-13}$. The next generation of atom-interferometry experiments currently ongoing should be able to achieve this goal, and together with an improved direct measurement, which is also in progress, could improve the sensitivity by another factor of $10$, only an order of magnitude away even from the
 benchmark~\eqref{naive_scaling}.

For the muon, the scaling with the lepton mass implies that less precision is required for the same BSM sensitivity, in the case of quadratic scaling by a factor $m_\mu^2/m_e^2\simeq 43\,000$, but only by $m_\mu/m_e\simeq 207$ in chirally enhanced scenarios, emphasizing the complementarity of $a_e$ and $a_\mu$ to not only resolve the flavor, but also the chirality structure of potential BSM contributions.
The current world average~\cite{\expref},
\begin{equation}
\label{eq:exp}
  \amuexp=\amuexpresult\times 10^{-11}\,,
\end{equation}
dominated by the FNAL E989 experiment, already lies within a factor of two of the benchmark~\eqref{naive_scaling}, and thus would allow one to exclude wide classes of BSM scenarios and parameter space corresponding to the myriad enhancements beyond the naive quadratic scaling at $\Lambda_\text{BSM}=1\TeV$.
Unfortunately, this sensitivity is currently significantly diluted by a larger uncertainty in the SM prediction~\cite{Aliberti:2025beg}
\begin{equation}
\label{eq:SM}
\amuSM=\amuSMresult\times 10^{-11}\,,
\end{equation}
as compiled by the Muon $g-2$ Theory Initiative based on Refs.~\cite{\SMref}. If this uncertainty could be reduced to the same level as Eq.~\eqref{eq:exp}, the resulting sensitivity, $|a_\mu^\text{BSM}|\lesssim 20\times 10^{-11}$, would reach the benchmark~\eqref{naive_scaling} up to a factor of three, testing scales
up to $\Lambda_\text{BSM}\simeq 600\GeV$, even in the most pessimistic case $c_L^\ell c_R^\ell R_\chi=1$.

\begin{figure}[t]
  \centering
\includegraphics[width=\columnwidth]{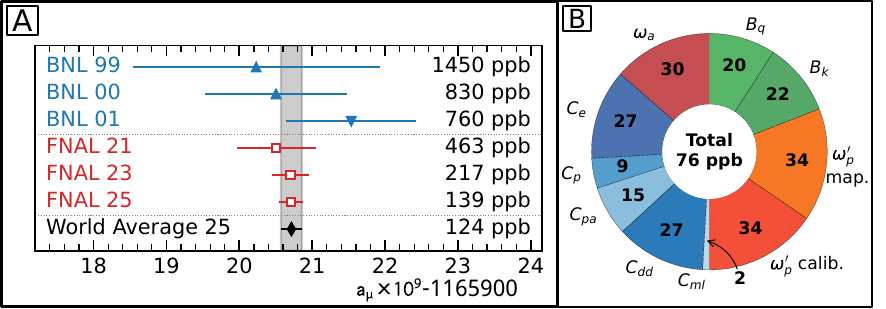}
  \caption{{\it A:} Experimental results from BNL (blue triangles)~\cite{Muong-2:2001kxu,Bennett:2002jb,Bennett:2004pv},
FNAL (red squares)~\cite{Muong-2:2021ojo,Muong-2:2023cdq,Muong-2:2025xyk}, and the world average  (black diamond). The uncertainties combine statistics and systematics. The quoted BNL $a_\mu$ values were computed by the FNAL Muon $g-2$ collaboration~\cite{Muong-2:2025xyk}, converting the published BNL-E821 results using updated values from CODATA 2022~\cite{Mohr:2024kco} for the external inputs of fundamental constants. BNL values are rounded to the nearest $10\ppb$.  All measurements employed positive muons except BNL-01, which ran with negative muons (inverted triangle). The CERN~III result~\cite{CERN-Mainz-Daresbury:1978ccd}, with a precision of $7\,300\ppb$, is not shown.  {\it B:} Final relatively balanced systematic uncertainties at FNAL, see Sec.~\ref{sec:FNAL} for a detailed discussion of the individual contributions.}
  \label{fig:world-average-pie-chart}
\end{figure}

In this review, we focus on the experimental techniques and innovations that led to Eq.~\eqref{eq:exp}. In particular, we focus on the evolution from the CERN $g-2$ experiment CERN-III
to the Brookhaven National Laboratory (BNL) experiment BNL-E821 and then to the FNAL
experiment FNAL-E989, as summarized in Fig.~\ref{fig:world-average-pie-chart}(A), see Secs.~\ref{sec:exp} and~\ref{sec:FNAL}. We then summarize
the current status of the SM prediction leading to Eq.~\eqref{eq:SM}, see Sec.~\eqref{sec:SM_prediction}. In Sec.~\ref{sec:future}, we describe ongoing efforts to improve the SM prediction to try and match the experimental precision, and contemplate future experiments that could surpass the current $124\ppb$.

\section{EXPERIMENTAL CONSIDERATIONS: WHY $g-2$ CAN BE MEASURED TO HIGH PRECISION}
\label{sec:exp}

Every modern\footnote{We define modern to imply the completed CERN-III, BNL-E821, and FNAL-E989 experiments.} muon $g-2$ experiment has relied on five ``miracles of nature'' that allow a measurement to be made with extremely high precision.
We present the basic experimental technique using short descriptions of each.

\subsection{Miracle \#1: The intrinsic properties of the muon}
The very existence of the muon is a miracle, the first particle discovered beyond the stable elements that make up our Universe.  Its intrinsic properties have led to critical studies of subatomic physics. The relatively long muon lifetime ($\simeq2.2\,\mu$s)\footnote{The lifetime has been measured to a precision of $1\ppm$, which determines the weak-interaction Fermi constant $G_{F}$ to $0.5\ppm$~\cite{MuLan:2012sih}.  The EW contribution $\amuEW$, see Eq.~\eqref{eq:EW_1loop} for the one-loop result, is expressed in terms of $G_F$ in the same scheme.} allows for the formation and transport of intense beams for detailed experimental investigations~\cite{Gorringe:2015cma}, including a precise measurement of its magnetic moment.

In 1956, Lee and Yang proposed that if the weak interaction violated parity, the decay of spin-$0$ pions would produce muons polarized along their direction of motion~\cite{Lee:1956qn}. Also, the angular distribution of positrons from the subsequent three-body $\mu^+ \to e^+\nu_e\bar{\nu}_\mu$ decay would be correlated with the direction of the muon spin.
Specifically, the differential decay distribution in the rest frame ($r$) is expressed as
\begin{equation}
\label{diff_decay}
    \frac{\odif{^2P}}{\odif{E}\odif{\cos\theta}} = N_r(E_r)[1+A_r(E_r)\cos\theta]\,,
\end{equation}
where $E_r$ is the positron energy and $\theta$ is the angle between its momentum and the muon spin.
The asymmetry function $A_r(E)$ represents the strength and sign of the correlation; the key will be that the highest-energy positrons are preferentially emitted in the direction of the positive muon's spin.
Both predictions were promptly confirmed~\cite{Garwin:1957hc,Friedman:1957}, demonstrating a violation of parity in the weak interaction and establishing the fundamental principles needed to measure the magnetic moment.

\subsection{Miracle \#2: The anomalous precession frequency is proportional to \amu}
All \gm\ experiments since the mid-1960s involve observing muons circulating in a magnetic storage ring (SR).
For a muon orbiting horizontally in the plane of a uniform vertical magnetic field $\vecB$, its cyclotron, $\vecomega_c$, and spin-precession, $\vecomega_s$, frequencies are given by
\begin{equation}
    \vecomega_{c} = -\frac{q}{m_\mu\gamma}\vecB\,, \qquad \vecomega_{s} = -g_\mu\frac{q}{2m_\mu}\vecB - (1-\gamma)\frac{q}{m_\mu\gamma}\vecB\,,
\end{equation}
where $\gamma=1/\sqrt{1-\vecbeta^2}$ is the Lorentz factor.
The difference between $\vecomega_s$ and $\vecomega_c$ is defined as the anomalous precession frequency,
\begin{equation}
\label{eq:precession}
    \vecomega_a \equiv \vecomega_s - \vecomega_c  = -\bigg(\frac{g_\mu-2}{2}\bigg)\frac{q}{m_\mu}\vecB \equiv -a_\mu \frac{q}{m_\mu}\vecB\,,
\end{equation}
which would vanish if $g_\mu$ were exactly equal to $2$.

Measurements of both $\omega_a=|\vecomega_a|$ and $B=|\vecB|$ yield a value for $a_\mu$.
We emphasize that this is a direct determination of the anomalous part of the magnetic moment,
thus gaining nearly three orders of magnitude in precision compared to  experiments performed in the rest frame, where $\omega_s \propto g_\mu$.
Equation~\eqref{eq:precession} embeds the equally critical fact that the relativistic $\gamma$ factor disappears in the final expression, which means that
a precise knowledge of the muon momentum is not required.

\subsection{Miracle \#3:  The magic $\gamma$}
To reach high precision, muons must be stored in the ring for approximately 10 lifetimes.
A system of  electrostatic quadrupole (ESQ) plates placed inside the SR provides weak vertical focusing.
The SR thus resembles a large Penning trap; see Fig.~\ref{fig:beamcollage}(E).
We use $\omega_a^m$ to denote the precession frequency actually measured in the experiment. This
differs from the nominal precession frequency because of the tiny influence of nonplanar trajectories and the presence of the electric field $\vecE$. A more complete\footnote{An additional perturbation to $\omega_a$ is required for a nonzero electric dipole moment, which has the form $\frac{\eta}{2}(\vecbeta \times\vecB + \vecE)$.} expression for \wa that includes these effects is given by~\cite{Bargmann:1959gz}
\begin{equation}
\vecomega_a^m =
- \frac{q}{m_\mu}\bigg[ \amu\vecB - a_{\mu}\bigg(\frac{\gamma}{\gamma+1}\bigg) (\vecbeta \cdot \vecB)\vecbeta
 - \bigg(\amu - \frac{1}{\gamma^2-1}\bigg)\vecbeta\times \vecE \bigg]\,.
   \label{eq:spin}
\end{equation}
The second term appears because of vertical betatron motion.\footnote{The spin equation \eqref{eq:precession} is derived assuming $\vecbeta \cdot \vecB = 0$, which is not exactly true in the SR.}
The third term accounts for the motional magnetic field felt by the relativistic muons in the presence of the electric field.
The CERN-III experiment~\cite{Bailey:1978mn} recognized that this term  vanishes for $\gamma=\sqrt{1+1/\amu} \simeq 29.3$ ($P_0 = 3.094\GeV, \gamma\tau_\mu \simeq 64.4\,\mu$s), which is called the ``magic momentum.''
All modern and completed experiments are based on this important realization.

A ``pitch'' correction ($C_p$), which depends on the average vertical amplitude distribution of the muons, and an electric field  correction ($C_e$), which depends on a measurement of the stored momentum distribution, are required to obtain the true $\omega_a$ from $\omega_a^m$.  These are both relatively large  adjustments compared to the final FNAL statistical uncertainty. We note that the planned J-PARC \gm\ experiment~\cite{Abe:2019thb} intends to inject lower-momentum, nonmagic muons with negligible transverse momentum into a compact ring that does not require electric quadrupole focusing.

\subsection{Miracle \#4:  The muon is its own polarimeter}
In the highly relativistic laboratory frame ($\gamma=29.3, \gamma\tau_\mu = 64.4\,\mu$s; $E_\text{max} = 3.1\GeV$), the positron energy is related to the energy and angular distribution in the rest frame to a good approximation by
\begin{equation}
E_\text{lab} \simeq \gamma E_{r}(1+\cos\theta),
\end{equation}
where $\theta$ is defined below Eq.~\eqref{diff_decay}.
The lab frame positron energy is strongly correlated to the muon spin direction.
A decay positron, having a momentum below $P_0$, curls to the inside of the SR where its energy ($E$) and time  ($t$) are recorded by a calorimeter.
The differential decay distribution has the form
\begin{equation}
\label{eq:wiggle}
     \frac{\odif{^2P}}{\odif{E}\odif{t}} \propto e^{-t/(\gamma\tau_\mu)} N(E)[1+A(E)\cos(\omega_a t - \phi(E))]\,,
\end{equation}
where $t$ is with respect to the injection time and $\phi(E)$ is a phase factor that represents the ensemble-averaged, and energy-dependent, spin orientation at $t = 0$.
These expressions illustrate that the polarimetry, meaning the average direction of the muon ensemble vs.\ time, is naturally correlated with the average energy distribution at that time of the decay positrons because of parity violation.
Equation~\eqref{eq:wiggle} describes the oscillatory plot shown in Fig.~\ref{fig:fft}(A).

\subsection{Miracle \#5:  The magnetic field co-magnetometers}
The final precision on \amu\  depends equally on the determination of \wa\ and $B$, see Eq.~\eqref{eq:precession}. More precisely, $B$ represents the average magnetic field seen by the muons whose decay positrons populate the final decay distribution fit. High-precision pulsed proton nuclear magnetic resonance (NMR) can determine the magnetic field strength to the few ppb level in small regions sampled by proton-rich probes containing pure water or petroleum jelly. A system of NMR probes, both fixed in space and scanned throughout the storage region, determines the entire SR field.
The key point is that the proton spin precesses in the same field as the muons; it is a co-magnetometer. 
From the ratio of the $\mu^+$ anomalous- to the $p$-precession frequencies, $\Rmu^{'}$,
one obtains \amu\ through the expression
\begin{equation}
a_{\mu} = \Rmu^{'}\frac{\mu'_{p}(\Tr)}{\mu^{}_{B}} \frac{m_{\mu}}{m_\text{e}}\,,
\label{eq:amueqnewcodata}
\end{equation}
where  $\mu'_p(\Tr)/\mu^{}_B$ is the ratio of the shielded proton magnetic moment to the Bohr magneton and $m_{\mu}/m_e$ is the muon-to-electron mass ratio~\cite{Mohr:2024kco}.
These well-known fundamental constants allow $\amu$ to be reevaluated if their values change in the future. 

\section{THE FERMILAB EXPERIMENT}
\label{sec:FNAL}
The FNAL-E989 experiment operated for six data-taking periods (Runs 1--6), all using positive muons.   The main elements of the BNL equipment reused at FNAL include the magnetic storage ring (SR)~\cite{Danby:2001eh}, the superconducting inflector magnet~\cite{Yamamoto:2002bb}, the vacuum chamber system, and the electrostatic quadrupole (ESQ) plates.
The fast kicker and ESQ power supplies are new, as are all 
detectors, electronic systems, precession and field data acquisition systems, and the pulsed-NMR probes and much of the electronics.   
The \gm\ results are published in three Physical Review Letters~\cite{Muong-2:2021ojo,Muong-2:2023cdq,Muong-2:2025xyk}.  Detailed reports on the methods to determine the precession frequency, beam dynamics, and magnetic field can be found in the Physical Review~\cite{Muong-2:2021vma,Muong-2:2021xzz,Muong-2:2021ovs,Muong-2:2024hpx}.

\subsection{Innovative aspects of the FNAL E989 experiment}
The modern SR experiments all follow the same experimental concept as described above, but each generation introduced significant improvements that resulted in a steady reduction in statistical and systematic uncertainties.
This section highlights a selection of unique and important aspects of the FNAL experiment, often following lessons learned from previous generations.  We also point out various limitations and unexpected problems that were encountered.

\begin{figure}[t]
	\includegraphics[width=1.\columnwidth]{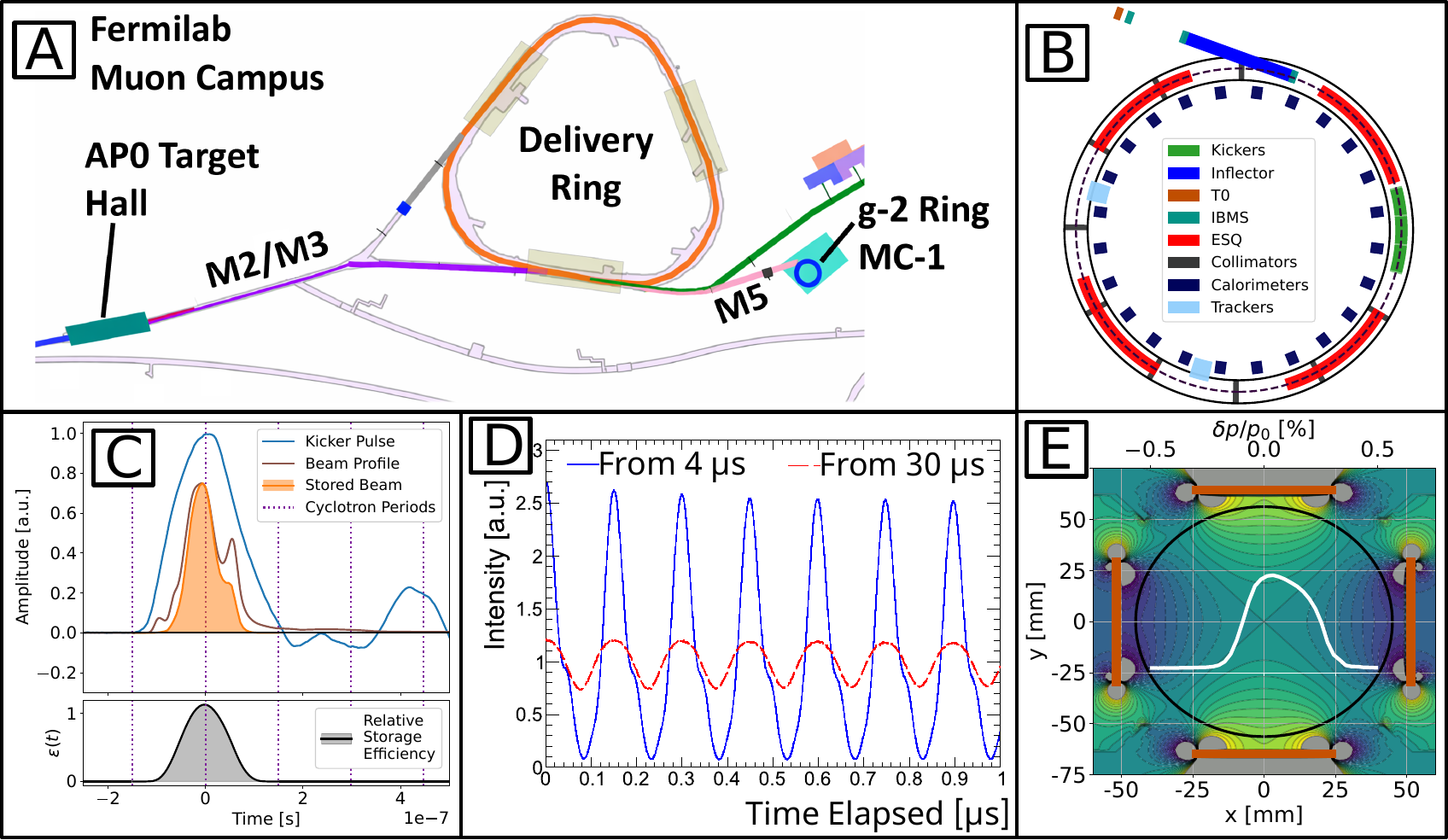}
	\caption{{\it A:} Muon Campus beamline overview and {\it B:} schematic of key elements in the SR. {\it C:} The entering beam intensity (brown) and kicker strength (blue) profiles vs.\ time of injection. The stored beam profile (solid orange) is the product of the $t_0$ profile and the overall storage efficiency (black, bottom) calculated from special systematic runs. {\it D:} The intensity of events striking a calorimeter at early (blue) and later (red) times. The peaks align with $\omega_c$ but the width evolves owing to the finite momentum spread in the bunch. Note, \wa and the exponential decay are divided out in this figure.  {\it E:} Momentum distribution (upper axis) of the stored beam overlaid with the ESQ plate positions and their electric field lines. The collimator is shown as a black ring.}
    \label{fig:beamcollage}
\end{figure}

\subsubsection{Injecting and storing polarized muons}
The most significant advance in each generation of experiments is arguably how muons are stored in the ring. 
In all cases, protons from an accelerator strike a target, producing $\simeq3.1\GeV$ pions that subsequently decay to the muons that are stored.
The pion decay length at this momentum is 174\,m.  CERN-III directly injected $\pi^+$ into their SR with the strategy that a small fraction (25\,ppm) will decay during the first turn in such a manner as to leave a $\mu^+$ on a storable orbit within the acceptance of the ring.
The majority of incoming pions produces a prompt hadronic ``flash'' seen by the detectors, and creates a slowly decaying background of thermal neutrons.  

BNL-E821 developed direct muon injection, an idea that works best if four conditions are met: 1) the upstream beamline collects and transports muons from pion decay in a bunch with temporal length $\ll T_c$, the cyclotron period; 2) undecayed pions and any protons in the bunch are removed upstream of the SR; 3) a field-free corridor is created in the SR to allow passage of the beam into the interior part of the ring; and 4) a fast magnetic kicker creates an outward $\simeq10$\,mrad angular deflection during the first turn to place the muons on the central orbit.
Most conditions at BNL were met, except 2). The relatively short pion-to-muon decay channel at BNL allowed a large flux of pions to enter the ring, which created a challenging background in the detector systems and ultimately affected the quality of the measurement.

At FNAL, the Booster injects intense proton pulses into the Recycler Ring, where a radio frequency (RF) system is used to form eight independent bunches, each with slightly different characteristic temporal shapes.
Bunches are extracted and made to strike a pion production target in the AP0 building; see Fig.~\ref{fig:beamcollage}(A) for this and the following descriptions. Positive pions (and protons) at $3.1\GeV$ are directed into the 279\,m-long M2/M3 beamline along which $\simeq80\%$ of $\pi^+$ decay to muons.
The mixture of $p$, $\pi^+$, and $\mu^+$ enters the 505\,m circumference Delivery Ring (DR), where it circulates four times before being deflected into the M5 beamline that ends at the entrance to the SR.
During these circulations, protons ($\beta = 0.94$) lag muons, and are swept away by a pulsed magnet; after four revolutions, essentially no pions remain.  This sequence produces a very pure $\simeq95\%$ polarized muon\footnote{Positrons from the target region are co-transported but rapidly vanish after a few turns due to synchrotron radiation energy loss.} injection, with a repetition frequency of $\simeq11$\,Hz.

The muon beam enters the SR through a hole in the back-leg of the magnet yoke, then passes  through a 1.7\,m-long narrow superconducting inflector magnet~\cite{Yamamoto:2002bb} that creates a field-free corridor into the storage region.
Figure~\ref{fig:beamcollage}(B) illustrates the locations of key components related to muon storage and the \wa measurement.
A newly developed pulsed magnetic kicker system (K1, K2, K3) provides  the needed angular deflection~\cite{Schreckenberger:2021kur}.  
Figure~\ref{fig:beamcollage}(C) illustrates the challenge of this process.
The vertical dashed lines are separated by the 149\,ns cyclotron period.
The kicker pulse strength (blue) is overlaid with a representative muon bunch intensity profile (brown).  The resulting stored profile (orange) is not uniform and differs slightly bunch to bunch.  The bottom panel illustrates the relative efficiency vs.\ time for this example.  Although the average of the pulse train can be nearly centered at the magic radius, $x=0$ in the figure, the relative storage efficiency varies from the beginning to the end of the bunch train.
Nevertheless, approximately 4000 muons are stored each ``fill'' with a momentum width $\delta P/P_0 \simeq 0.15\%$, and a mean momentum nearly centered on the magic momentum.

The 9\,cm  diameter storage region is defined by the collimators (black ring in Fig.~\ref{fig:beamcollage}(E)). 
The ESQ field lines are shown in the figure.
One subtle improvement was modifying the high-voltage (HV) standoffs and replacing the aluminum outer plate with thin aluminized Mylar in the ESQ region immediately downstream of the inflector. This reduced the material through which muons must first pass before entering the storage volume and increased the storage fraction.

The 32 individual plates that make up the ESQ system are raised to HV just prior to each fill.  In two of four regions, some plates are charged in two HV steps, such that the muon bunch is first displaced with respect to the central radius before being centered a few $\mu$s later.  The particles having trajectories that might eventually strike a collimator are removed by this two-step procedure, which minimizes ``muon loss'' when the precession frequency data are evaluated.

\begin{figure}[t]
	\includegraphics[width=\columnwidth]{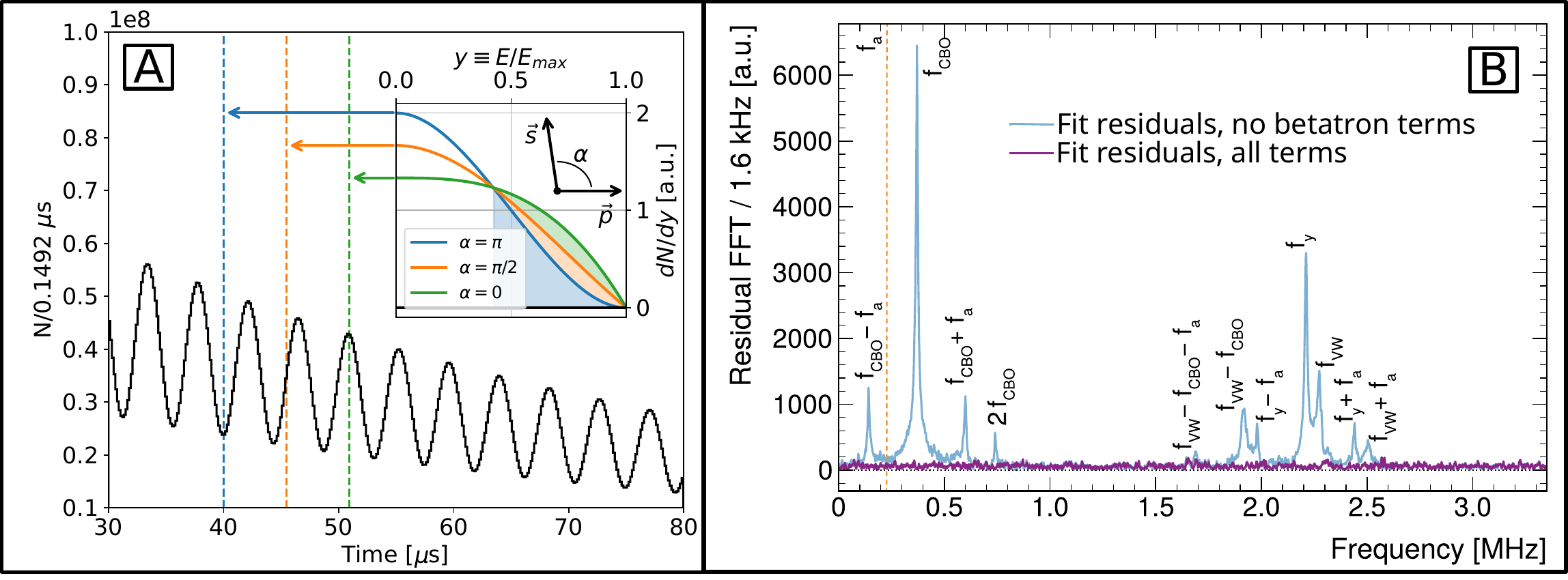}
	\caption{{\it A:} A zoom-in on a portion of the precession data plot with an inset to show the relationship of spin and momentum vectors to the peaks and troughs of the data. {\it B:} The light blue trace shows the FFT of the residuals to a 5-parameter fit of the positron decay spectrum $N(t)$. The red dashed line stands at the $\omega_a$ frequency, which is well fit.  The largest peaks are associated with CBO and the vertical betatron motion.  The flat purple trace is the same procedure applied after additional terms are included in the fit function.}
    \label{fig:fft}
\end{figure}

\subsubsection{Measuring the precession frequency}
All decay positrons curl to the inside of the SR, where their time-of-arrival and energy are recorded in one of the 24 electromagnetic calorimeters symmetrically positioned around the ring~\cite{Fienberg:2014kka,Kaspar:2016ofv,Muong-2:2019beb}.  Each calorimeter contains 54 lead-fluoride Cherenkov crystals read out by individual silicon photo-multipliers, and arranged in a 6 high by 9 wide matrix.  The segmented geometry and few-ns pulse duration is critical to minimize pileup, a dominant systematic at BNL but a negligible systematic uncertainty in this experiment.

The positrons having higher energies form a spectrum $N(t)$ as shown in Fig.~\ref{fig:fft}(A).
A fit to $N(t)$ determines $\omega_a^{m}$, Eq.~\eqref{eq:spin}.
Events entering the spectrum are weighted by an effective, measured energy-dependent asymmetry, $A(E)$, which varies smoothly from $0.0$ to $0.8$ across the 1.0 to 3.1\,GeV interval of included positron energies.   
An innovative system~\cite{Muong-2:2019hxt} of pulsed lasers flashes each crystal in patterns to establish long- and short-term gain corrections functions that are applied during reconstruction of the raw data. 
The stability of the energy response is therefore maintained throughout a fill to a few parts in $10^{-4}$ to avoid time dependence in $\phi(E)$.

The blue curve in Fig.~\ref{fig:fft}(B) shows the fast Fourier transform (FFT) of the residuals of a fit of $N(t)$ using the five-parameter function $N_{0}e^{-t/(\gamma\tau_\mu)}[1+A\cos(\omega_{a}t-\phi)]$.
The sharp peaks indicate modulations of $N(t)$ caused by betatron oscillations.
The mean of the stored muon ensemble executes harmonic motion in the horizontal and vertical planes and its width expands and contracts.
The frequencies are well-determined from the field index $n = (\kappa R_0)/(\beta B_0)$, where $R_0$ is the ring radius, and $\kappa$ is the electric quadrupole gradient that is dependent on the ESQ HV settings.
The horizontal and vertical betatron frequencies are given by $f_x = f_c\sqrt{1-n}$ and $f_y = f_c\sqrt{n}$, respectively, where $f_c = \omega_c/(2\pi)$ and $n \simeq 0.107$.
Because $f_x$ exceeds the Nyquist limit of $f_c/2$, an alias frequency $f_\text{CBO} = f_c - f_x$ is imprinted on the individual detectors.  This coherent betatron oscillation (CBO) frequency is the strongest peak in the FFT plot.

The measured precession frequency $\omega_a^{m}$ is shifted from the true $\omega_a$ by $\simeq 800$\,ppb  and the $\chi^2$ is unacceptable if CBO is ignored in the fitting function.\footnote{In Runs-5/6, an RF system applied a carefully timed HV on the ESQ plates, which reduced the CBO amplitude 
by a factor of 10.}
Additional terms are needed in the fit function to account for the peaks appearing in the FFT plot.  More challenging is determining the functions that describe their decoherence vs.\ time in the fill.  A good fit typically involves $\Order{(40)}$ parameters to achieve a $\chi^2$  consistent with 1.0 and a flat FFT residual spectrum as seen in Fig.~\ref{fig:fft} in purple.

In the final stages of the Run-4/5/6 analysis, a potential sub-$10^{-4}$ gain instability was evaluated for pairs of events occurring in the same detector a few $\mu$s apart in time, an ``intermediate''  interval that had not been fully explored. The frequency of events of this type diminishes with the muon lifetime. A tiny effect was identified in post-run lab studies leading to a correction procedure that was applied to the Run-4/5/6 data sets prior to unblinding. It was also used to update the previously published results from Runs 1--3.  Details are given in Ref.~\cite{Muong-2:2025xyk}.

\subsubsection{Beam diagnostics instrumentation}
In a \gm\ experiment,  muons are not directly measured by detectors once they are in the ring. However, where they are stored and how they move within the ring is critically important to correcting \wam and to determining the magnetic field experienced by the muons.  We describe instrumentation that determines the intensity and spatial profile of the incoming beam, and the dynamics of the stored muon ensemble.

The injected beam intensity vs.\ time is measured by the $T0$ counter (orange trace in Fig.~\ref{fig:beamcollage}(C)). The fill-by-fill transverse profile is measured just outside of the SR and also immediately upstream of the inflector magnet by arrays of scintillating fibers (IBMS).   This information is used to optimize beam tuning and monitor fill-by-fill stability.
The transverse profile of the stored muons inside the ring is measured directly, but only in special runs that are not included in the final data set.  The minimally intrusive scintillating fiber (MiniSciFi) system is inserted near $180^{\circ}$ or $270^{\circ}$ and overlaps the beam during storage.  The transverse profile vs.\ time-in-fill is measured by scanning three 0.5\,mm diameter fibers across the aperture.   A second detector determines the vertical distributions using rotated fibers.  In both cases, the impact on the beam is minor, and an accurate map of the mean and width is obtained.

Two in-vacuum straw tracker systems~\cite{King:2021hst} continuously monitor the stored beam.
Each station indirectly creates an image of the stored beam vs.\ time by reconstructing the trajectories of positrons that curl inward back to their decay locations.  This yields an excellent estimate of the muon coordinates at the time of decay.  The tracker information determines the vertical amplitude distribution for the pitch correction and informs all beam dynamics corrections and SR modeling simulations.

The initial injection of muons is bunched in time such that decay positrons seen by an individual calorimeter are aligned with the cyclotron frequency, as shown in Fig.~\ref{fig:beamcollage}(D) for early (blue) and later (red) times following injection.  The evolution in time is caused by the finite momentum spread which causes muons to orbit the ring at slightly different periods.  Unfolding this debunching sequence determines the momentum distribution that is stored, which is illustrated by the white trace in Fig.~\ref{fig:beamcollage}(E); top axis. 
This momentum mean value and width are critical to establishing the electric-field correction, $C_e$.

An example of the importance of measuring the muon profile vs.\ time-in-fill that was not originally anticipated occurred during Run-1.
A fault in 2 of the 32 HV resistors in the ESQ system led to long RC time constants to charge those plates.  This caused the mean and width of the muon ensemble to move systematically during the fill.  Understanding the implications led to the discovery of large correlations between the transverse decay position within the ring and the average spin phase; i.e.,  $\phi_{(x,y)}$, where $\phi$ refers to the phase term in Eq.~\eqref{eq:wiggle}. 
All muons precess at the same frequency, even if their individual spin phases at $t=0$ and average transverse decay locations differ.  If the distribution is stable, the average of these phases is a constant.  

A time-dependent profile leads to a time-dependent average phase, $\phi \rightarrow \phi(t)$, which manifests itself as a perturbation in \wa.  To leading order, the argument of the cosine term $(\omega_a t + \phi(t))$ in Eq.~\eqref{eq:wiggle} becomes $(\omega't + \phi)$  and thus $\omega' \neq \omega_a$.
Even after repairs fixed the resistor problem, a much smaller residual motion persists but is well understood.

To summarize, corrections to the measured precession frequency \wam\ are necessary because of the electric field ($C_e$), pitch ($C_p$), muon loss ($C_{ml}$), and ``phase-acceptance'' ($C_{pa}$) effects. An additional differential decay ($C_{dd}$) correction is largely driven by the need to accommodate the time-dependent average momentum distribution shown in the lower panel of Fig.~\ref{fig:beamcollage}(C).

\subsubsection{Measuring the magnetic field}
The magnetic field $\vecB$ in Eq.~\eqref{eq:spin} is measured using pulsed proton NMR techniques and is expressed as $\opprimetildeatTexp$, which is the proton spin precession frequency inside a room-temperature spherical water sample and averaged over the stored muon distribution.

\begin{figure}[t]
	\includegraphics[width=1.\columnwidth]{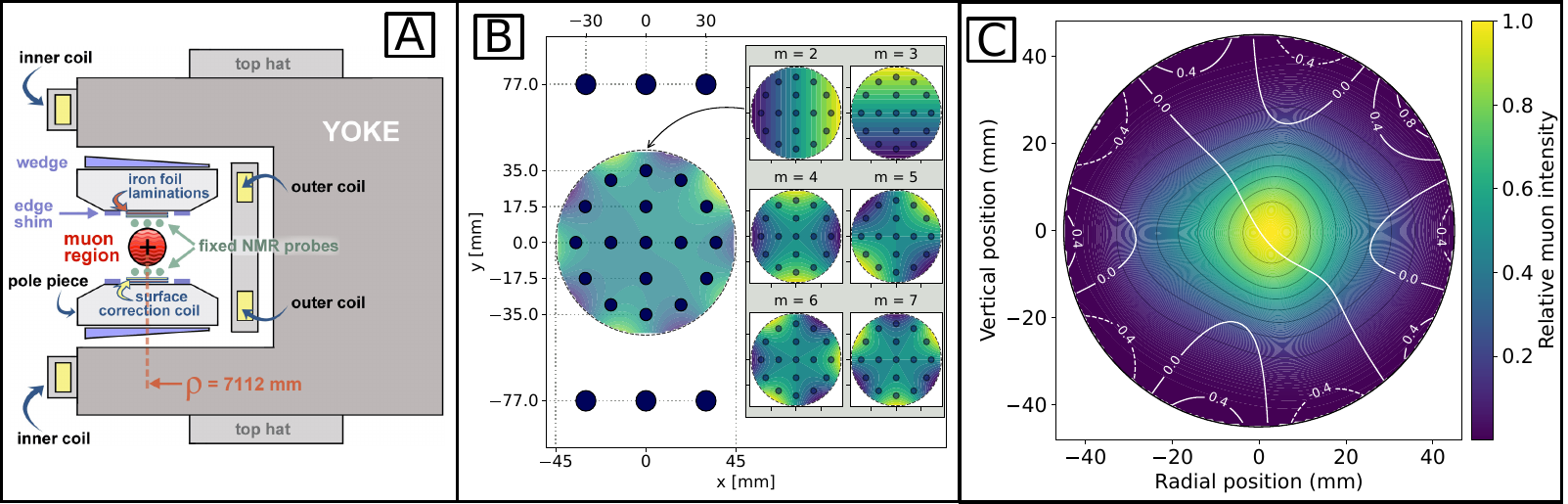}
	\caption{{\it A:} Cross section of the C-shaped SR magnet illustrating the physical adjustments to develop a uniform field and the locations of the fixed NMR probes above and below the muon storage region. {\it B:} The NMR trolley probe locations, above which is a station with six fixed probes that are embedded just outside of the vacuum chamber.  The small figures represent the multipoles of the field beyond the dipole that are determined by combinations of the NMR probes.  {\it C:} The azimuthally averaged field contours and the stored muon relative intensity from Run-5.}
    \label{fig:fieldcollage}
\end{figure}

A special-purpose highly uniform magnetic-resonance-imaging (MRI) magnet was used to develop a custom high-purity water~\cite{Flay:2021jrs} probe to determine the absolute magnetic field~\cite{Muong-2:2021ovs}. The probe was cross-checked with a He-3 probe~\cite{Farooq:2020swf} and used as part of an in-ring movable ``plunging'' probe system, which in turn cross-calibrates one-by-one the 17 probes in an NMR trolley~\cite{Corrodi:2020sav}.
Figure~\ref{fig:fieldcollage}(B) shows the placement of these probes on the front face of the trolley. The insets indicate how combinations of these probes are used to determine the moment distribution of the field.
This device periodically maps the entire SR field {\it in situ}. The temporal variation of the field is continuously monitored using an additional set of 378 probes~\cite{Swanson:2024xnr} placed above and below the storage region, as shown in the figure.

Improving the uniformity of the magnetic field is key to reducing systematic uncertainties. After reassembling and shimming the SR magnet at FNAL using the existing tools shown in Fig.~\ref{fig:fieldcollage}(A), the average dipole field was no better than what had been achieved at BNL.
Detailed mapping of this field using a 25-element shimming trolley provided the needed data to develop a solution in which thousands of small iron foil laminations were patterned on plates that were inserted in the air gap between the pole pieces and glued to the flat surface.  This had the effect of fine-tuning the average field as a function of azimuth and controlling gradients in the direction
transverse to the beam propagation.  The overall uniformity of the field improved by more than a factor of two.

The final field average is made from hundreds of trolley maps acquired during the runs. Each trolley map provided the local magnetic field strength at 150,000 positions in the storage ring. Changes in the field between trolley maps were monitored with a set of 378 fixed NMR probes outside the storage volume. Each map is weighted by the number of positron events obtained in the interval between maps. Folding the field maps with the stored muon distribution, described by a moment distribution $M$, gives the needed field experienced by the muons and is expressed as $\langle \opprimeatTexp\times M\rangle$.
A representative example is shown in Fig.~\ref{fig:fieldcollage}(C).

An anticipated magnetic transient $B_k$ occurs when the fast-kicker fires and produces an eddy current in the vacuum chambers. Its persistence into the early part of the fill causes an effect on the average field.  It is measured using two independent magnetometers based on the Faraday rotation of polarized light through terbium gallium garnet (TGG) crystals.

An unexpected transient $B_q$ was discovered that is associated with energizing the quadrupole plates prior to each injection cycle.  This caused the plates to vibrate and create a slowly oscillating magnetic field that had to be carefully mapped in time to determine its residual effect during the muon storage time.

\subsection{The FNAL E989 results}
The frequency ratio \Rmuprime, corrected for beam dynamics and field transient effects, is given by 
\begin{equation} \label{eq:R}
   \mathcal{R}_\mu' = \frac{\wam \big(1+C^{}_{e}+C^{}_{p}+C^{}_{pa}+C^{}_{dd}+C^{}_{m l}\big)}{\langle \opprimeatTexp\times M\rangle(1+B^{}_k+B^{}_q )}\,.
\end{equation}
Equation~\eqref{eq:amueqnewcodata} connects $\mathcal{R}_\mu'$ to \amu\ and the fundamental constants.

\begin{table}[t]
\caption{Corrections and uncertainties from combined FNAL running periods, and current uncertainties due to the external parameters.  Numerical summary courtesy of A.~Lusiani.}
\label{tab:uncertainties}
\begin{tabular}{l r r}\hline
Quantity & Correction (ppb) & Uncertainty (ppb)\\
\hline
\wam\ (statistical) & -- & $98$\\
Beam dynamic (BD) $C_i$ corrections to \wam to yield \wa & $+538$ & $43$ \\
Net \wa\ (statistical + BD + fitting uncertainties )  & -- & $111$\\
\hline
\opprime\ (all systematics) & --  & $48$\\
Magnetic field transient $B_i$ corrections  to \opprime\ & $-58$ & $31$ \\
Net \opprime\ uncertainty  & -- & $57$\\
\hline
Total uncertainties for \Rmuprime & -- & $125$\\
\hline
$\mu'_p/\mu_B$ and $m_\mu/m_e$ external parameters & -- & $23$ \\
\hline
Total for \amu\  &  & $127$ \\\hline
\end{tabular}
\end{table}

The true values for \wam\ and \opprime remain blinded during each publication campaign cycle until all corrections have been established and the fits of data by multiple independent teams  are consistent.
More than 300 billion decay positrons result in a statistical uncertainty of $98\ppb$.  The combined precession and field systematics total is $76\ppb$.
Large corrections from the electric field and pitch effects, combined with additional smaller ones, shift the raw value of \Rmu by a factor $\simeq4$ times larger than the final uncertainty, see Table~\ref{tab:uncertainties} and Fig.~\ref{fig:world-average-pie-chart}(B). Each correction is studied in extreme detail by multiple internally blinded teams, giving confidence in the adjustments.  However, in any future experiment, one would aim to reduce the size of these corrections.

\section{STATUS OF THE STANDARD-MODEL PREDICTION}
\label{sec:SM_prediction}

\begin{figure}[t]
    \includegraphics[width=\linewidth]{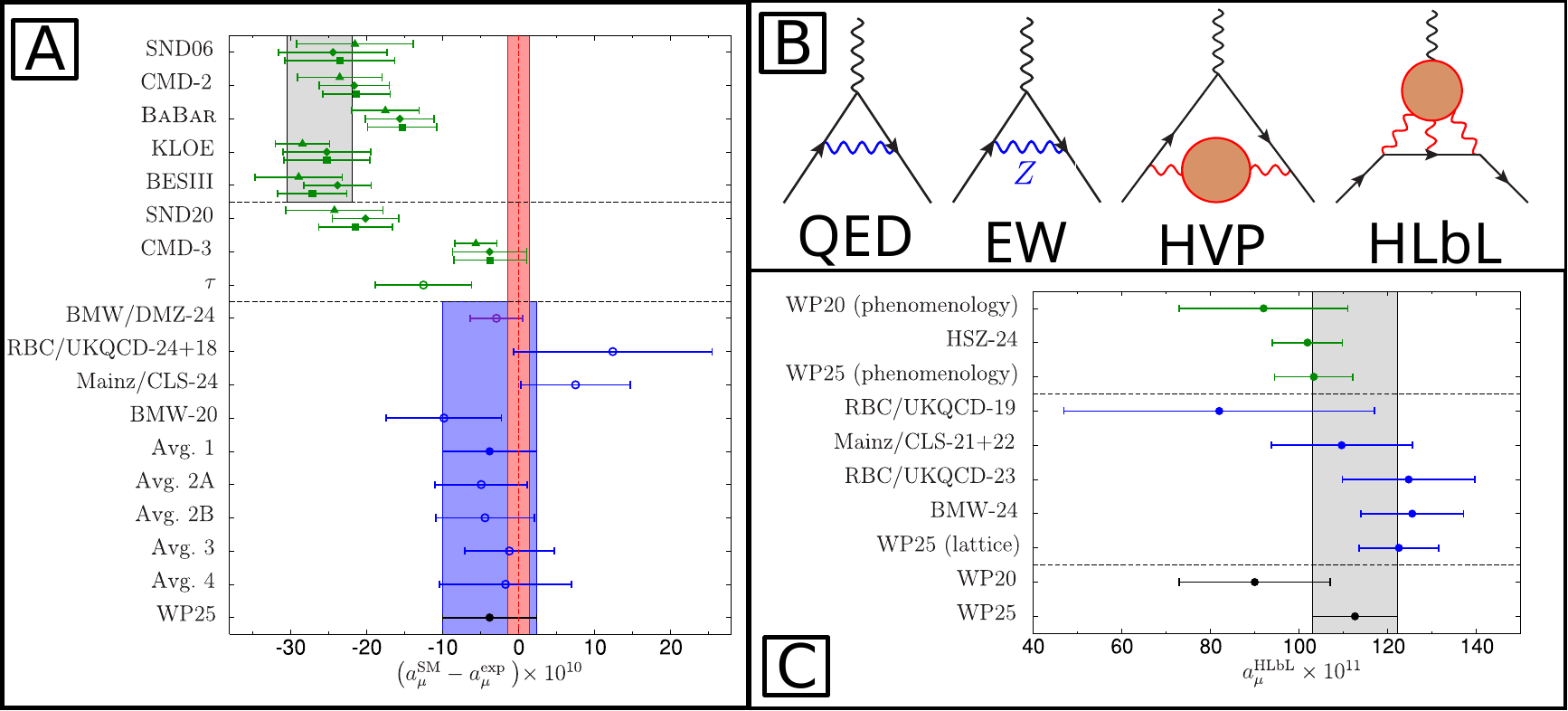}
\caption{{\it A:} Summary of various determinations of $\amuHVPLO$, propagated to $\amuSM$. The first two panels refer to data-driven determinations, where the three points for each $e^+e^-$ experiment reflect the CHKLS~\cite{Colangelo:2018mtw,Colangelo:2022prz,Stoffer:2023gba,Leplumey:2025kvv} (triangle), DHMZ~\cite{Davier:2017zfy,Davier:2019can,Davier:2023fpl} (diamond), and KNTW~\cite{Keshavarzi:2018mgv, Keshavarzi:2019abf,Keshavarzi:2024wow} (square) methods, see WP25 for details. The gray band indicates the WP20~\cite{Aoyama:2020ynm} result, based on the $e^+e^-$ experiments above the first dashed line.  The second panel represents $e^+e^-$ experiments that became available afterwards as well as the $\tau$-based estimate from WP25, based on Refs.~\cite{Davier:2023fpl,Davier:2010fmf,Castro:2024prg,Colangelo:2022prz,Hoferichter:2023sli}.
The last panel summarizes lattice-QCD determinations, including the hybrid evaluation by BMW/DMZ-24~\cite{Boccaletti:2024guq}, three  individual lattice-QCD calculations by RBC/UKQCD-24+18~\cite{RBC:2024fic},  Mainz/CLS-24~\cite{Djukanovic:2024cmq}, BMW-20~\cite{Borsanyi:2020mff},  and five lattice HVP averages from WP25. The blue band refers to the final WP25 result, which coincides with ``Avg.~1.''
In all cases, except for the gray WP20 band, the remaining contributions to $\amuSM$  beyond $\amuHVPLO$ are taken from WP25.
The red band denotes the experimental world average. Adapted from WP25.
{\it B:} Sample diagrams for $a_\mu$ in the SM: solid lines refer to the muon, wiggly lines to photons, and the red blobs to hadronic matrix elements. The first diagram gives the leading QED contribution by Schwinger~\cite{Schwinger:1948iu}, the second one represents a one-loop EW diagram with $Z$ exchange, and the last two diagrams correspond to HVP and HLbL topologies, respectively.
{\it C:} Summary of HLbL evaluations, from data-driven methods (green), lattice QCD (blue), and combinations (black). The averages are from WP20 and WP25, respectively, the other points refer to HSZ-24~\cite{Hoferichter:2024vbu,Hoferichter:2024bae}, RBC/UKQCD-19~\cite{Blum:2019ugy}, Mainz/CLS-21+22~\cite{Chao:2021tvp,Chao:2022xzg}, RBC/UKQCD-23~\cite{Blum:2023vlm}, and BMW-24~\cite{Fodor:2024jyn}. Adapted from WP25.}
    \label{fig:SM_prediction}
\end{figure}

The SM prediction of $a_\mu$ decomposes into pure quantum electrodynamics (QED) contributions, electroweak (EW) effects, and hadronic corrections
\begin{equation}
 \amuSM=\amuQED+\amuEW+\amuhad\,,\qquad \amuhad=\amuHVP+\amuHLbL\,,
\end{equation}
where the latter materialize from quark loops, which are, in principle, described by quantum chromodynamics (QCD), but at low energies require nonperturbative techniques because the quarks and gluons are confined into hadrons. They
are further separated into topologies featuring hadronic vacuum polarization (HVP) and hadronic light-by-light scattering (HLbL), see Fig.~\ref{fig:SM_prediction}(B) for representative diagrams in each category. Predicting $\amuSM$ at the level $\Order(10^{-10})$ requires exquisite control over all parts of the SM, and it is quite remarkable that this is indeed possible.
In the following,
we discuss the status for each class of corrections. Our discussion is
strongly based on the White Papers (WP) of the Muon $g-2$ Theory Initiative (\path{https://muon-gm2-theory.illinois.edu/}), to which we will refer
as WP20~\cite{Aoyama:2020ynm} and WP25~\cite{Aliberti:2025beg} throughout this section, each reflecting a community consensus at the respective time. The resulting bottom line is summarized in Table~\ref{tab:summary}.

\begin{table}[t]
\caption{Summary of all contributions to the SM prediction and comparison to experiment, adapted from WP25.
}
\label{tab:summary}
\begin{center}
\begin{tabular}{l r l}\hline
	   Contribution &  Value   $\times 10^{11}$ & References\\ \hline
	   Experiment (E989, E821)  & $\amuexpresult$ & Refs.~\cite{\expref}  \\\hline
HVP LO (lattice) & $\amuHVPLOresult$ & Refs.~\cite{\latticeHVPref}\\
\textcolor{gray}{HVP LO ($e^+e^-,\tau$)}  & \multicolumn{2}{l}{\textcolor{gray}{Estimates not provided at this point}}\\
HVP NLO ($e^+e^-$) &  $\amuHVPNLOresult$ & Refs.~\cite{Keshavarzi:2019abf,DiLuzio:2024sps}\\
HVP NNLO ($e^+e^-$) &  $\amuHVPNNLOresult$ & Ref.~\cite{Kurz:2014wya} \\
HLbL (phenomenology) & $\amuHLbLdataresult$ & Refs.~\cite{\dataHLbLref}\\
HLbL NLO (phenomenology) &  $\amuHLbLNLOdataresult$ & Ref.~\cite{Colangelo:2014qya}\\
HLbL (lattice)  & $\amuHLbLlatticeresult$ & Refs.~\cite{\latticeHLbLref}\\
HLbL (phenomenology + lattice) & $\amuHLbLaverageresult$ & Refs.~\cite{\dataHLbLref,\latticeHLbLref}\\
\hline
QED               &  $\amuQEDresult$  & Refs.~\cite{\QEDref}\\
EW      & $\amuEWresult$   & Refs.~\cite{\EWref}\\
HVP LO (lattice) + HVP N(N)LO ($e^+e^-$)  & $\amuHVPtotalresult$ &  Refs.~\cite{\HVPref}\\
HLbL (phenomenology + lattice + NLO)  & $\amuHLbLtotalresult$ & Refs.~\cite{\HLbLref}\\
      Total SM Value   & $\amuSMresult$  & Refs.~\cite{\SMref}\\
      Difference:    $\Delta a_\mu\equiv\amuexp - \amuSM$  & $\amudiffresult$ & \\\hline
\end{tabular}
\end{center}
\end{table}

\subsection{QED and electroweak contributions}

By far the dominant contribution to $\amuSM$ arises from QED, including all diagrams with purely photonic and leptonic ($\ell=e,\mu,\tau$) loops.
Starting from Schwinger's seminal result~\cite{Schwinger:1948iu}, the expansion can be expressed as
\begin{align}
\label{eq:QED}
 \amuQED&=\frac{\alpha}{2\pi}+0.765\,857\,420(8)_e(11)_\tau[13]\bigg(\frac{\alpha}{\pi}\bigg)^2
 +24.050\,509\,77(17)_e(16)_\tau[23]\bigg(\frac{\alpha}{\pi}\bigg)^3\notag\\
 &+130.8782(60)_\text{MC}\bigg(\frac{\alpha}{\pi}\bigg)^4
 +750.2(9)_\text{MC}\bigg(\frac{\alpha}{\pi}\bigg)^5+\Order\big(\alpha^6\big)\,,
\end{align}
where the uncertainties derive from $m_\mu/m_e=206.7682827(46)$, $m_\tau/m_\mu=16.8170(11)$~\cite{Mohr:2024kco}, and Monte-Carlo (MC) integration, respectively.
The complete five-loop QED result~\eqref{eq:QED} represents the crowning achievement of decades of precision QED calculations.
Up to $\Order(\alpha^3)$, analytical results are available, both for the mass-independent terms~\cite{Petermann:1957hs,Sommerfield:1958,Laporta:1996mq} and mass-dependent corrections~\cite{Elend:1966a,Laporta:1992pa,Laporta:1993ju}, so that the uncertainties in the coefficients in Eq.~\eqref{eq:QED} only depend on the input for the lepton mass ratios. At $\Order(\alpha^4)$, the mass-independent term is known essentially analytically~\cite{Laporta:2017okg}, while for the mass-dependent corrections the complexity of the calculation no longer allows for analytical solutions and the loop integrals need to be performed by MC integration, whose error dominates the respective uncertainty budget~\cite{Kinoshita:2005zr,Aoyama:2012wk}. Finally, at $\Order(\alpha^5)$ both mass-independent and mass-dependent contributions are evaluated numerically, and the size of the mass-independent term recently decreased by $\simeq 0.8$ upon improved MC sampling~\cite{Aoyama:2012wk,Volkov:2019phy,Volkov:2024yzc,Aoyama:2024aly}, about the same size as the MC uncertainty in the mass-dependent correction. Numerically, the largest contributions to the coefficients arise from terms enhanced by powers of $\log\frac{m_\mu}{m_e}$ due to electron loops (and factors of $\pi^2$, originating from electron light-by-light topologies), explaining their rapid increase.
For the evaluation of $\amuQED$, input for the fine-structure constant $\alpha$ is required, the current most precise determinations being~\cite{Aliberti:2025beg}
\begin{align}
\alpha^{-1}[\text{Cs}] &= 137.035\,999\,045(27)\,,     \label{eq:alpha_Cs} \\
\alpha^{-1}[\text{Rb}] &=  137.035\,999\, 2052(97)\, ,      \label{eq:alpha_Rb}\\
\alpha^{-1}[a_e]  &=  137.035\,999\,163(15)\,,
\label{eq:alpha_ae}
\end{align}
derived from atom-interferometry measurements in Cs~\cite{Parker:2018vye}, Rb~\cite{Morel:2020dww}, and the anomalous magnetic moment of the electron $a_e$~\cite{Fan:2022eto}, respectively. The tension between $\alpha^{-1}[\text{Cs}]$ and $\alpha^{-1}[\text{Rb}]$, already alluded to in Sec.~\ref{sec:intro}, translates to an uncertainty of about $0.14\times 10^{-11}$ in $\amuQED$. This is the same level at which $\Order(\alpha^6)$ contributions are expected to enter, estimated based on the maximally possible enhancement with $\log\frac{m_\mu}{m_e}$ at six-loop order~\cite{Aoyama:2012wk}. The combination of these two effects leads to the final uncertainty quoted for $\amuQED$ in Table~\ref{tab:summary}.

Loop contributions that involve $W$, $Z$, or Higgs bosons are summarized in the EW contribution $\amuEW$. In fact, prior to the Higgs discovery, lack of the knowledge of the Higgs boson mass constituted the largest uncertainty in $\amuEW$.
At one-loop order, the result can be expressed as~\cite{Jackiw:1972jz,Bars:1972pe,Altarelli:1972nc,Bardeen:1972vi,Fujikawa:1972fe}
\begin{equation}
\label{eq:EW_1loop}
\amuEW[1\text{-loop}] =
  \frac{G_{F}}{\sqrt{2}}
  \frac{m _\mu ^{2}}{8 \pi ^2}
  \bigg[
    \frac{5}{3}
    +
    \frac{1}{3}(1-4 s_{\text{W}} ^2)^2+\Order\bigg(\frac{m_\mu^2}{M_{W}^2},\frac{m_\mu^2}{M_H^2}\bigg)
    \bigg]=194.79(1)\times 10^{-11}\,,
\end{equation}
with on-shell weak mixing angle $s_W^2=\sin^2\theta_W=1-M_W^2/M_Z^2$ and the SM prediction expressed in terms of the Fermi constant $G_F$ as measured in muon decay~\cite{MuLan:2012sih}, $M_Z$, and $\alpha$, while $M_W$ is a derived quantity~\cite{Gnendiger:2013pva}. Two-loop contributions are sizable, $\amuEW[2\text{-loop}]=-40.4(4)\times 10^{-11}$, and are typically further separated into  bosonic~\cite{Czarnecki:1995sz,Heinemeyer:2004yq,Gribouk:2005ee}, Higgs-dependent~\cite{Czarnecki:1995wq,Gnendiger:2013pva}, and other fermionic contributions~\cite{Peris:1995bb,Knecht:2002hr,Czarnecki:2002nt}. In particular, the latter include quark triangle loops, for which nonperturbative effects become important. Improvements in their evaluation~\cite{Ludtke:2024ase,Hoferichter:2025yih} together with the consideration of $\alpha_s$ corrections for the heavy quarks~\cite{Melnikov:2006qb} and lattice-QCD input for $\gamma$--$Z$ mixing~\cite{Ce:2022eix} lead to the updated result~\cite{Hoferichter:2025yih}
\begin{equation}
\amuEW=\amuEWresult\times 10^{-11}\,,
\end{equation}
where three-loop effects are estimated to be $\lesssim 0.2\times 10^{-11}$~\cite{Czarnecki:2002nt,Degrassi:1998es}.

\subsection{Hadronic light-by-light scattering}
\label{sec:HLbL}

HLbL contributions enter at $\Order(\alpha^3)$, involving the hadronic four-point function as indicated in Fig.~\ref{fig:SM_prediction}(B).
For a long time this contribution was considered the theoretically most challenging one, to the extent that concerns about whether HLbL scattering could be put onto solid grounds might have played a role in the decision to terminate the BNL experiment.
This situation changed thanks to both the development of dispersive techniques~\cite{Colangelo:2014dfa,Colangelo:2014pva,Colangelo:2015ama} and methodological advances in lattice QCD~\cite{Blum:2014oka,Blum:2017cer,Asmussen:2022oql}. The former 
allow for a more direct connection to data for $\gamma^{(*)}\gamma^{(*)}\to\text{hadrons}$, which constrain the poles and cuts of the HLbL tensor and can thus be used to reconstruct the full amplitude via dispersion relations. The latter have allowed one to control the challenging MC sampling, including an important contribution from quark-disconnected diagrams, and the extrapolation to infinite volume. These developments have led to two independent HLbL determinations at a level of precision around $10\%$, see Ref.~\cite{Aliberti:2025beg} for a detailed review.

A summary of the present situation is shown in Fig.~\ref{fig:SM_prediction}(C), including phenomenological~\cite{\dataHLbLref} and lattice-QCD~\cite{\latticeHLbLref} evaluations. They agree  at a level of $1.5\sigma$, leading to the overall average
\begin{equation}
\label{HLbL_summary}
    \amuHLbL[\text{phenomenology + lattice}]=\amuHLbLaverageresult\times 10^{-11}\,,
\end{equation}
whose error includes a scale factor $S=1.5$.
This comparison emphasizes the benefit of having two independent approaches available, as the validation provides further confidence in the final result~\eqref{HLbL_summary}, while the mild tension motivates further improvements in either method in the future.
Finally, adding an estimate for next-to-leading-order (NLO) HLbL topologies~\cite{Colangelo:2014qya} produces the final HLbL number quoted in Table~\ref{tab:summary}.

\subsection{Hadronic vacuum polarization}
\label{sec:HVP}

The leading, and most critical, hadronic effect already arises at $\Order(\alpha^2)$ in the form of the HVP diagram shown in Fig.~\ref{fig:SM_prediction}(B).
A priori, it is not at all evident that such a complicated nonperturbative object can be evaluated with sub-percent precision, and
for a long time HVP evaluations have therefore relied exclusively on the master formula~\cite{Bouchiat:1961lbg,Brodsky:1967sr}
\begin{equation}
\label{g-2:eq:amu_HVP_master}
    \amuHVPLO=\bigg(\frac{\alpha m_\mu}{3\pi}\bigg)^2\int_{s_\text{thr}}^\infty ds \frac{\hat K(s)}{s^2}R_\text{had}(s)\,,
\end{equation}
which allows one to express its effect via a dispersion integral in terms of the hadronic $R$-ratio,
\begin{equation}
\label{g-2:eq:R_ratio}
R_\text{had}(s)=\frac{3s}{4\pi\alpha^2}\sigma\big[e^+e^-\to\text{hadrons}(+\gamma)\big] \,,
\end{equation}
and a kernel function $\hat K(s)$, establishing a direct link to experimental data for hadronic cross sections.
At the required level of precision, it is important
to be precise about the QED effects included in this quantity. $R_\text{had}(s)$ does not include
QED corrections to the photon propagator, but it does include final states with photons in
addition to hadrons. It also includes effects of virtual photons coupling to quarks, but not
virtual photons also coupling to the initial state $e^+e^-$.  Accordingly, the integration in Eq.~\eqref{g-2:eq:amu_HVP_master} starts at  the threshold of the $e^+e^-\to\pi^0\gamma$ channel, $s_\text{thr}=M_{\pi^0}^2$. In WP20, the final result for the HVP contribution was based on $e^+e^-$ data, including data for the by far numerically dominant $e^+e^-\to\pi^+\pi^-$ channel from the SND06~\cite{Achasov:2006vp}, CMD-2~\cite{CMD-2:2006gxt}, \babar~\cite{BaBar:2012bdw}, KLOE~\cite{KLOE-2:2017fda}, and BESIII~\cite{BESIII:2015equ} experiments. Moreover, different methodologies for the compilation of $e^+e^-$ data were taken into account~\cite{Davier:2017zfy,Keshavarzi:2018mgv,Colangelo:2018mtw,Hoferichter:2019mqg,Davier:2019can,Keshavarzi:2019abf}, and an additional uncertainty beyond a simple scale factor was introduced to account for the tension between the \babar{} and KLOE measurements, leading to the gray band in Fig.~\ref{fig:SM_prediction}(A).
Subsequently, new $\pi^+\pi^-$ measurements from SND20~\cite{SND:2020nwa} and CMD-3~\cite{CMD-3:2023alj,CMD-3:2023rfe} have become available, and while the former falls within the previous range, the latter increases tensions to a level that cannot be taken into account anymore by a meaningful inflation of uncertainties. Due to these severe tensions, combined with the fact that most experiments do not cover the entire phase space, already different methodologies to assign HVP integrals for a given $\pi^+\pi^-$ experiment can lead to different results, as illustrated by the triangle, diamond, and square points in Fig.~\ref{fig:SM_prediction}(A).

An alternative data-driven evaluation relies on $\tau\to\nu_\tau+\text{hadrons}$ decays~\cite{Alemany:1997tn}, in which case isospin symmetry provides access to the isovector part of the hadronic cross section as long as isospin-breaking corrections~\cite{Cirigliano:2001er,Cirigliano:2002pv,Flores-Baez:2006yiq,Davier:2010fmf,Miranda:2020wdg,Castro:2024prg} can be controlled. A critical review of these corrections is provided in WP25, identifying several effects that precluded the use of $\tau$ data for the HVP evaluation due to a remaining model dependence that is difficult to quantify in a reliable manner. However, the best estimate from WP25, based on Refs.~\cite{Davier:2023fpl,Davier:2010fmf,Castro:2024prg,Colangelo:2022prz,Hoferichter:2023sli}, is included in the second panel of Fig.~\ref{fig:SM_prediction}(A).

While the HVP evaluation from data-driven methods is thus inconclusive at this point, material progress has been achieved since WP20 in lattice QCD, based on the time-momentum representation of the hadronic two-point function~\cite{Bernecker:2011gh}. In particular, detailed cross checks have been performed using windows in Euclidean time~\cite{RBC:2018dos} as a tool to scrutinize the different systematic errors of the calculation. In total, the final average quoted in WP25,
\begin{equation}
\label{eq:HVP}
 \amuHVPLO[\text{lattice}]=\amuHVPLOresult\times 10^{-11}\,,
\end{equation}
is based on 17 different papers from 8 independent lattice-QCD collaborations \cite{ExtendedTwistedMass:2024nyi, MILC:2024ryz, Spiegel:2024dec, Boccaletti:2024guq, Kuberski:2024bcj, ExtendedTwistedMass:2022jpw, Wang:2022lkq, RBC:2023pvn, Borsanyi:2020mff, Ce:2022kxy, Aubin:2022hgm, Lehner:2020crt, RBC:2018dos, Djukanovic:2024cmq, FermilabLatticeHPQCD:2024ppc, RBC:2024fic, Giusti:2019xct}, from which three almost complete
lattice calculations of the entire LO HVP contribution are derived~\cite{Borsanyi:2020mff,RBC:2024fic,Djukanovic:2024cmq}, see Fig.~\ref{fig:SM_prediction}(A). Moreover, different strategies to combine the individual results, corresponding to the five averages shown in Fig.~\ref{fig:SM_prediction}(A), prove very stable, and additional systematic errors have been assigned to cover
the most challenging aspects of the calculation related to the evaluation of isospin-breaking effects and the noisy contributions at long distances.

\subsection{Status in 2025}

The status of the SM prediction $\amuSM$, as reproduced in Table~\ref{tab:summary}, reflects a snapshot of the situation as in 2025. In particular, at this point in time
$e^+e^-$ data are not used in WP25 for the LO HVP evaluation, until the origin of the discrepancies among data sets is better understood.\footnote{The exception concerns higher-order HVP iterations~\cite{Calmet:1976kd,Kurz:2014wya}, for which the increased uncertainty can be tolerated~\cite{Keshavarzi:2019abf,DiLuzio:2024sps}. Mixed leptonic and hadronic corrections at $\Order(\alpha^4)$ are $\lesssim 1\times 10^{-11}$~\cite{Hoferichter:2021wyj}.} The new substantial tensions after the CMD-3 measurement, rising above the $5\sigma$ level in comparison to KLOE, simply preclude a meaningful average right now, justifying the major change compared to WP20. To the contrary, not including data from hadronic $\tau$ decays until better-controlled evaluations of the isospin-breaking corrections become available is in line with the previous WP20 procedure.

In the meantime, the consolidation of lattice-QCD results for HVP justifies
the use of Eq.~\eqref{eq:HVP} for the final
SM prediction in WP25,\footnote{This result agrees with the hybrid evaluation by BMW/DMZ-24~\cite{Boccaletti:2024guq}, using data input for the long-distance tail.} thus reversing the situation as in WP20. We emphasize again that this is only a representation of the situation at present, and it is fully expected that data-driven evaluations will again enter the SM prediction in the future, thanks to the ongoing efforts described in Sec.~\ref{sec:future}. In fact, the importance of having two independent methods cannot be overstated, as exemplified by the evaluation of HLbL scattering, and a resolution of the current puzzles in HVP is motivated not just by $\amuSM$, but also by HVP contributions to $a_e$, the hadronic running of $\alpha$ and the weak mixing angle, and muonium hyperfine structure~\cite{DiLuzio:2024sps}.

\section{GOING FORWARD}
\label{sec:future}

\subsection{Prospects for a Standard-Model prediction at 124 ppb}

Matching the precision of $124\ppb$ achieved in experiment, Eq.~\eqref{eq:exp}, poses a challenge to theory, Eq.~\eqref{eq:SM}, which currently lags behind by about a factor of four. However, to fully leverage the BSM sensitivity such improvements are absolutely critical, as described in Sec.~\ref{sec:intro}. Here, we summarize ongoing efforts towards this goal~\cite{Colangelo:2022jxc}, focusing on the crucial HVP contribution.\footnote{Also the HLbL contribution will receive further scrutiny over the next years, anticipating improvements both from data-driven evaluations and lattice QCD, to better understand the mild tension discussed in Sec.~\ref{sec:HLbL} and further improve the precision.}

\nocite{Polat:2026ysh,Campanario:2019mjh,Ignatov:2022iou,Colangelo:2022lzg,Abbiendi:2022liz,BaBar:2023xiy,Budassi:2024whw,Aliberti:2024fpq,Fang:2025mhn,Budassi:2026lmr}

\begin{figure}[t]
\includegraphics[width=\linewidth]{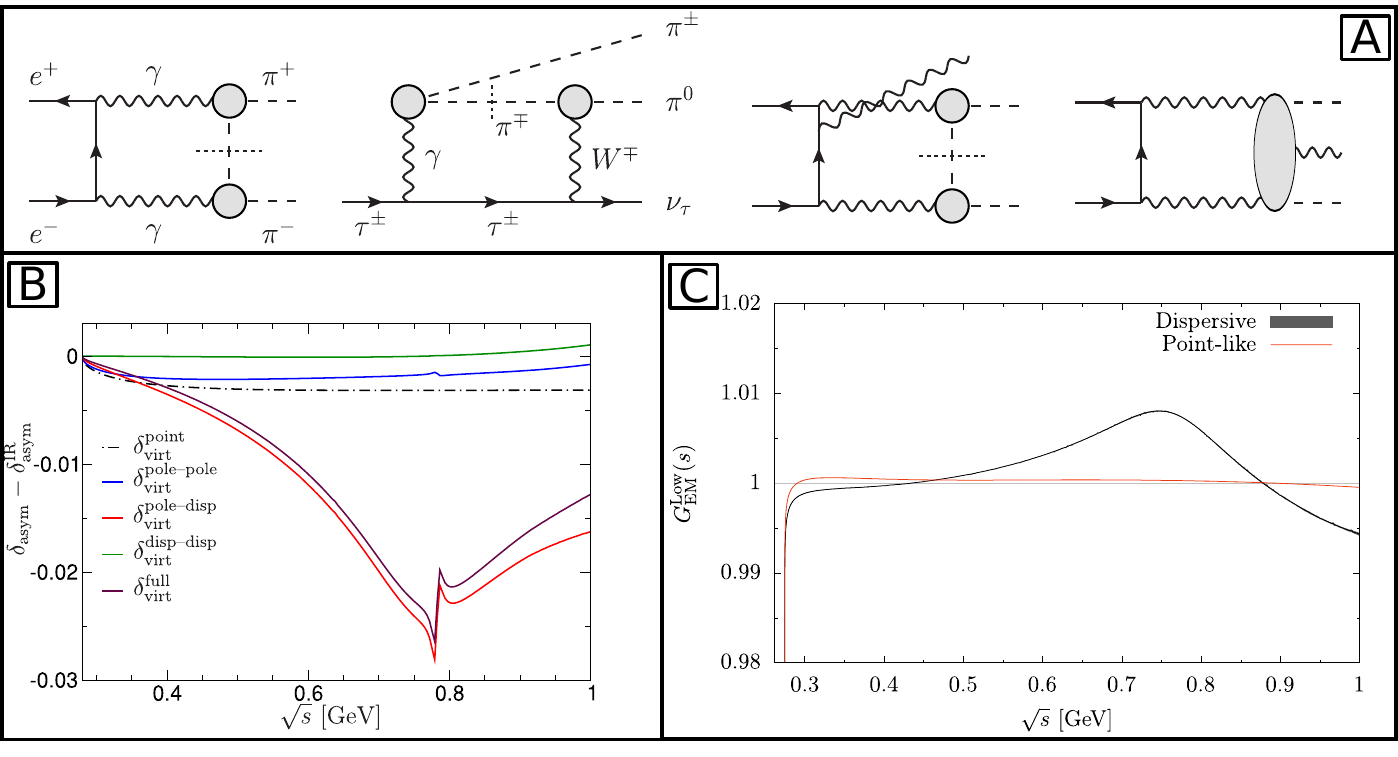}
\caption{{\it A:} Structure-dependent radiative corrections to the forward--backward asymmetry in $e^+e^-\to\pi^+\pi^-$, $\tau^\pm\to\pi^\pm\pi^0\nu_\tau$ decays, and $e^+e^-\to\pi^+\pi^-\gamma$, respectively. The short-dashed lines denote pions taken on-shell, the gray circles refer to the pion vector form factor, and the gray oval indicates the general $\gamma^*\gamma^*\gamma\to\pi^+\pi^-$ matrix element. {\it B:} Virtual contributions to the forward--backward asymmetry $\delta_\text{asym}$ in $e^+e^-\to\pi^+\pi^-$, with the infrared (IR) divergence subtracted. The black dot-dashed line indicates the point-like approximation, while the solid maroon line gives the full dispersive result, broken down into pole--pole (blue), pole--dispersive (red), and dispersive--dispersive (green) components. Adapted from Ref.~\cite{Colangelo:2022lzg}.
{\it C:} Radiative correction factor $G_\text{EM}$ for $\tau^\pm \to\pi^\pm\pi^0\nu_\tau$ in leading Low approximation. The red curve corresponds to the point-like limit, the black curve to the full dispersive result. Adapted from Refs.~\cite{Colangelo:2025iad,Colangelo:2025ivq}.}
    \label{fig:theory_prospects}
\end{figure}

\subsubsection{$e^+e^-\to\pi^+\pi^-(\gamma)$}

An important aspect of resolving the puzzling situation in $e^+e^-\to\pi^+\pi^-$, see Fig.~\ref{fig:SM_prediction}(A), involves new data and analyses. In the meantime, such new results have become available from \babar, based on data that had never been analyzed before, and SND. Preliminary results~\cite{Polat:2026ysh} suggest that the new \babar{} analysis almost coincides with their previous result, while the new results from SND move closer to CMD-3 compared to their earlier measurement~\cite{SND:2020nwa}.
Similarly to \babar, a large amount of data from the KLOE experiment had never been analyzed, and an effort to do so is ongoing. Finally, new data are expected from BESIII and  Belle II.

In all cases, special emphasis is put on the role of radiative corrections and MC generators~\cite{Campanario:2019mjh,Ignatov:2022iou,Colangelo:2022lzg,Abbiendi:2022liz,BaBar:2023xiy,Budassi:2024whw,Aliberti:2024fpq,Fang:2025mhn,Budassi:2026lmr}, to scrutinize whether relevant effects were missed in previous analyses and ensure state-of-the-art input for ongoing ones.
In fact, one of the cross checks performed by CMD-3~\cite{CMD-3:2023alj,CMD-3:2023rfe} concerns the forward--backward asymmetry in $e^+e^-\to\pi^+\pi^-$, which is $C$-odd, thus canceling in the total cross section, but still constitutes an interesting test case that could have implications for $C$-even contributions in measurements using initial-state radiation, see Fig.~\ref{fig:theory_prospects}(A). It was observed that the MC simulation based on scalar QED (sQED) for the pion states could not reproduce the data, a mismatch that could be resolved by structure-dependent radiative corrections~\cite{Ignatov:2022iou}, as shown in Fig.~\ref{fig:theory_prospects}(A). In particular, the first diagram contributes to the asymmetry, interfering with the tree-level amplitude, and is only poorly represented by the point-like, sQED approximation. In the language of dispersion relations~\cite{Colangelo:2022lzg}, this topology involves the general pion Compton tensor~\cite{Garcia-Martin:2010kyn,Hoferichter:2011wk,Moussallam:2013una,Danilkin:2018qfn,Hoferichter:2019nlq,Danilkin:2019opj}, but the leading contribution originates from the pion pole. In the end, it is the interplay of pole and dispersive corrections, as a remnant of the cancellation of the IR singularity, that displays an enhancement with the $\rho(770)$ resonance, see Fig.~\ref{fig:theory_prospects}(B). This resonance effect causes a sizable enhancement beyond the point-like limit, explaining the observation by CMD-3.

In their case, there are no direct consequences for the cross section $e^+e^-\to\pi^+\pi^-$, since the asymmetry vanishes upon integration over the entire phase space due to $C$ parity. In contrast, for experiments using the initial-state-radiation technique---\babar, Belle II, BESIII, KLOE---$C$-even contributions involving similar topologies exist, see the last two diagrams in Fig.~\ref{fig:theory_prospects}(A). The first one should again be reasonably well described by the pion pole, at least at sufficiently low energies, while the second diagram, in principle, requires the full $\gamma^*\gamma^*\gamma\to\pi^+\pi^-$ matrix element as input. Both topologies are currently under scrutiny in the {\it RadioMonteCarLow 2} effort (\path{https://radiomontecarlow2.gitlab.io/}), to implement both effects in MC generators, for the latter using a simplified prescription that at least captures the pion form factors within the loop integral, while more complete amplitudes for $\gamma^*\gamma^*\gamma\to\pi^+\pi^-$ are being developed. Given the different center-of-mass energies and experimental cuts it is well conceivable that such corrections could lead to a better line-up of the different $e^+e^-\to\pi^+\pi^-$ experiments, at least, CMD-3 demonstrated that sizable corrections beyond the point-like approximation are possible, and a similar observation was recently made in the context of hadronic $\tau$ decays~\cite{Colangelo:2025iad,Colangelo:2025ivq}, to which we turn next.

\subsubsection{Hadronic $\tau$ decays}

In Ref.~\cite{Aliberti:2025beg} three main effects were identified that cast doubts on the reliability of the currently available isospin-breaking corrections. First, the matching of short-distance corrections with hadronic matrix elements involved an $\Order(\alpha/\pi)$ scheme ambiguity~\cite{Cirigliano:2023fnz},
which has now been addressed~\cite{Cirigliano:2026ios}
using input from lattice QCD for the nonperturbative matching~\cite{Feng:2020zdc,Yoo:2023gln}. Second,  resonance-enhanced virtual corrections could become important, see second diagram in Fig.~\ref{fig:theory_prospects}(A), and indeed these effects were evaluated recently~\cite{Colangelo:2025iad,Colangelo:2025ivq}, again leading to sizable corrections beyond the point-like limit, see Fig.~\ref{fig:theory_prospects}(C). Third, isospin breaking in the matrix elements needs to be quantified, which constitutes the most challenging task, but could be addressed using lattice QCD~\cite{Bruno:2018ono} and dispersive techniques~\cite{Colangelo:2025iuq}.
In addition to these crucial theory advances, also new data especially on the $\tau\to\pi\pi\nu_\tau$ spectrum would be extremely valuable, as could become available at Belle II~\cite{Belle-II:2018jsg}.

\subsubsection{Lattice QCD}

To improve the precision of lattice-QCD calculations of HVP, as a first step the differences among the present calculations are being investigated---which become most significant when using the pion decay constant for the scale setting---including the conversion to a common isospin scheme as suggested in Ref.~\cite{FlavourLatticeAveragingGroupFLAG:2024oxs}. To move towards $124\ppb$ in precision, the most pressing challenges concern 
the statistics of the noisy long-distance tail (both isospin-limit and QED corrections),  performing more and improved calculations of isospin-breaking corrections, and controlling the continuum extrapolation at a level that is commensurate with the goal of few permil precision;  see, e.g., Refs.~\cite{FermilabLattice:2025dui,Lehner:2025qrl} for recent work addressing scale setting and noise reduction. Crucially, sustained funding of the necessary computational resources is needed to reach the precision goal set by the FNAL E989 experiment.

\subsubsection{MUonE}

The MUonE experiment~\cite{CarloniCalame:2015obs,MUonE:2016hru} aims at a measurement of the HVP contribution in muon--electron scattering via the space-like hadronic running of $\alpha$, $\Delta\alpha_\text{had}(t)$, yielding another independent determination with completely different systematics. A 2025 test run aims at a precision of $20\%$ (statistical and systematics each) in $\Delta\alpha_\text{had}(t)$, while a full-scale proposal will be developed during the CERN Long Shutdown 3. Apart from the experimental challenges, also MC generators at $10\ppm$ need to be developed~\cite{Banerjee:2020tdt}.

\subsubsection{Towards 124 ppb precision}

Reaching $124\ppb$ precision, or $\Delta \amuSM=14.5\times 10^{-11}$, will require a combination of methods. Prior to CMD-3, determinations based on $e^+e^-$ data had reached an uncertainty of $24\times 10^{-11}$~\cite{Keshavarzi:2019abf}, without assigning additional systematic effects for the \babar{}--KLOE tension, so this level appears realistic if the tensions among the $e^+e^-\to\pi^+\pi^-$ data sets can be resolved. The current experimental uncertainty propagated from hadronic $\tau$ decays amounts to $28\times 10^{-11}$~\cite{Aliberti:2025beg}, so if improved isospin-breaking corrections and new data from Belle II become available, hadronic $\tau$ decays may contribute at a similar level of precision as $e^+e^-$ data. Improving current lattice-QCD calculations by a factor of two, as realistic for the near-term future, corresponds to about $0.5\%$ precision or $35\times 10^{-11}$ for a single calculation, and further improvements are possible by combining results from different collaborations, improving the attainable statistical error by another factor of two. Finally, if lattice-QCD and data-driven evaluations agree, the precision can be further improved via combined evaluations, profiting from each method's advantages in different parts of the HVP integral~\cite{RBC:2018dos,Boccaletti:2024guq}. To this end, also detailed comparisons of lattice-QCD and data-driven evaluations will be critical~\cite{Colangelo:2022vok,Benton:2023dci,Hoferichter:2023sli,Davier:2023cyp,Benton:2024kwp}.

\subsection{Can an experiment go beyond 124 ppb?}

The history of $g-2$ experiments is long and marked by a series of innovative advances, each enabling a new generation of measurements with ever increasing precision. Although the current experimental uncertainty of $124\ppb$ now exceeds that of the SM prediction, this situation may not persist indefinitely, motivating consideration of how the measurement might ultimately be pushed even further. One such effort, the J-PARC E34 experiment, remains under development and therefore still has opportunities to incorporate design features aimed at achieving the next advance in sensitivity. More speculatively, one may also ask whether the FNAL E989 technique itself could be extended.

\subsubsection{The J-PARC E34 Experiment}

The J-PARC E34 experiment~\cite{Abe:2019thb} breaks the mold of continuing the lineage of magic momentum \gm\ experiments.  A comparison of key parameters between E34 and E989 is found in Ref.~\cite{Gorringe:2015cma}, see also Ref.~\cite{Zhang:2025gmt} for another recent proposal.  Briefly, by creating a pure muon beam with negligible transverse momentum, one can run at any convenient momentum and avoid the need for electric quadrupole focusing.  The plan of E34 is to create a source of stopped muonium atoms, strip them with a laser system, and accelerate them from rest using RFQ technology to a momentum of $\simeq300\MeV$. This beam is injected into a highly uniform ``MRI'' sized magnet following a spiraling trajectory.  It is softly kicked into a horizontal plane, in which the muons orbit.  The decay positrons will be collected internally by a ring of tracking detectors.  This challenging experiment features many new ideas that are being developed.  A recent success has been the stopping, stripping, and reacceleration of muons to an energy so far of $100\keV$~\cite{Aritome:2024rlu}.

The original $\simeq450\ppb$ precision goal of E34 is not competitive with FNAL, but the collaboration continues to evaluate possible improvements and aims to match the FNAL precision. The main value is the very different technique that would yield an independent validation.

\subsubsection{At FNAL}

Inevitably, the question arises: \emph{Can one do better at FNAL?}
Despite the present limitations of the SM prediction, we describe a thought exercise that illustrates a potential path toward an improved experimental precision by roughly a factor of three---to the level of $\simeq40\ppb$---should future developments warrant it.\footnote{The ideas presented here have contributions from T.~Barrett, S.~Foster, D.~Kawall, J.~LaBounty, B.~Morse, Y.~Semertzidis, and V.~Tishchenko and we emphasize that each needs a dedicated study.}
The concept envisions a ten-fold increase in statistics ($30\ppb$) combined with a three-fold reduction in systematics ($25\ppb$).
The external constant uncertainties should decrease to about $10\ppb$ because of anticipated future muonium experiments~\cite{MuSEUM:2025cmo}.

In this discussion, the existing FNAL SR, beamlines, and supporting infrastructure are largely retained.
The PIP-II LINAC upgrade~\cite{2969223} is expected to deliver a $30\%$ higher proton flux at a $33\%$ faster repetition cycle.
It will be necessary to upgrade the Recycler RF system to better rebunch the injected proton batches into shorter pulses.
This will both improve the kicker efficiency and largely eliminate the systematic uncertainty from differential decay (see Fig.~\ref{fig:beamcollage}(C)).
Current \gm\ simulations indicate that 25\,ns-long bunches are stored 1.7 times more efficiently and another multiplicative factor of 1.3 can be realized by installing the built, but never used, open-ended inflector.
Additional nearly two-fold storage efficiency is expected if the ESQ voltages can be raised to their design values, a challenge that remains owing to sparks that limited the maximum voltage in E989.
These upgrades collectively increase the number of stored muons per fill.
In E989, four of 21 proton batches in the accelerator supercycle were allocated to \gm, with the remainder serving concurrent neutrino experiments.
Redirecting the full supercycle to a new \gm\ run could yield up to a five-fold increase in data accumulation.
Increasing the density (spacing) of the quadrupole magnets in the M2 FODO lattice would also result in a higher muon rate per pulse. 
Combined with the other improvements, a (10--20) fold increase in statistics appears realistic---though certain necessary changes described next reduce the storage.

Figure~\ref{fig:world-average-pie-chart}(B) shows the relatively balanced final systematic uncertainties from E989.
The $C_i$ terms correspond to the uncertainties on the calculated corrections that are made to $\omega_a^m$ prior to unblinding.
The corrections and their uncertainties depend in various degrees on the transverse size, transverse motion, momentum width, and the nonideal kicker-pulse overlap as shown in Fig.~\ref{fig:beamcollage}(C).
Centering the beam and reducing the storage aperture can substantially suppress detector acceptance variations, minimize betatron oscillations, and reduce the variance of phase vs.\ decay coordinate.
The electric field correction, $C_e$, depends on knowledge of the mean and width of the momentum distribution, which is determined by observing how the incoming muon bunch evolves in time. Its uncertainty would be reduced if the overlap of the kicker and incoming pulse were better optimized.  The size of the correction would be reduced if a smaller momentum width were obtained, which follows from a smaller aperture.   Simulations using a 4-cm aperture suggest a reduction by a factor of two in the size of $C_e$.

The uncertainty driving the differential decay correction $C_{dd}$ is also related to the nonideal kick of the incoming bunch, as discussed earlier.
The largest systematic uncertainties associated with the fit to extract $\omega_a^m$   arise from CBO and vertical motion.
Both effects can be reduced with a full deployment of the RF damping system that was introduced in the later stages of E989 and the much smaller storage region.

The $B_q$ correction to the magnetic field can be reduced or eliminated by engineering a more mechanically stable ESQ plate system to avoid the vibrations observed during HV ramping process that induced a residual oscillatory magnetic field.
The $B_k$ contribution, related to eddy currents in the kicker plates and vacuum chambers, can be reduced via improved structural design and through dedicated beam-off studies focusing on the smaller storage region.
The determination of the average magnetic field experienced by the muons is also advantaged here, both by allowing field probes to be placed closer to the beam and the fact that the field uniformity is better in the smaller, centered region. 
Absolute field calibration improvement is a challenge.  Ideas floated include use of $^3$He magnetometers for both the calibration reference and, if feasible, within the trolley.
These probes offer reduced sensitivity to gradients, temperature, and geometry-dependent effects compared to the current NMR system.
Finally, improved thermal stability within the experimental hall is critical for both magnetic field and detector stability.

\subsection{Final thoughts}

For years to come, the standard for $a_\mu$ as a precision probe of physics beyond the SM will be set by the FNAL E989 experiment, having reached a precision of $127\ppb$ as the result of a dedicated experimental campaign that started in 2018 with Run 1 and concluded with the announcement of the final results of Runs 1--6 in 2025. Over the next years, it will thus be the task of theory to catch up, and hopefully match the experimental precision to fully leverage the BSM sensitivity. A number of developments are ongoing towards this goal, including new data for the critical $e^+e^-\to\pi^+\pi^-$ channel, improved radiative corrections and MC generators, lattice-QCD calculations at increased precision, improved isospin-breaking corrections and new data for hadronic $\tau$ decays, as well as the MUonE experiment. At the same time, the J-PARC E34 experiment strives to match E989's precision with a new method, to provide independent validation of the experimental result. At the end of this program, we expect a conclusive SM test at the level of precision pioneered by E989, and time will tell if future developments will allow one to go even beyond.

%Disclosure
\section*{DISCLOSURE STATEMENT}
The authors are not aware of any affiliations, memberships, funding, or financial holdings that might be perceived as affecting the objectivity of this review. 

% Acknowledgements
\section*{ACKNOWLEDGMENTS}
Financial support by the Swiss National Science Foundation Project No.\ TMCG-2\_213690 (MH) and the United States Department of Energy grant DE-FG02-97ER41020 (DH) is gratefully acknowledged.
The authors wish to thank M.~Incagli, D.~Kawall,  J.~LaBounty, P.~Winter, and H.~Wittig for reviewing and comments. Custom figures were supplied by S.~Holz and J.~LaBounty.

% References

\bibliographystyle{ar-style5.bst}
\bibliography{g-2}

\end{document}